\documentclass[usenatbib]{mn2e}

\usepackage{graphicx}
\usepackage{mathptmx}
\usepackage{subfigure}
\usepackage{amssymb}
\usepackage{amsmath}
\graphicspath{{./Images/}}
\begin{document}

\title[Obscured star formation at $z={\it 0.84}$]{Obscured star
  formation at z = 0.84 with HiZELS\thanks{Based on observations
  obtained with the Wide Field CAMera (WFCAM) on the United Kingdom
  Infrared Telescope (UKIRT), as part of the `High-$z$ Emission Line
  Survey' (HiZELS) campaign project.}: the relationship between star
  formation rate and H$\alpha$ or ultra-violet dust extinction}

\author[T.\ Garn et al.]{Timothy Garn$^{1}$\thanks{E-mail:
    tsg@roe.ac.uk}, David Sobral$^{1}$, Philip N.\ Best$^{1}$, James
    E.\ Geach$^{2}$, Ian Smail$^{2}$, 
\newauthor Michele Cirasuolo$^{3}$, Gavin B.\ Dalton$^{4,5}$, James
    S.\ Dunlop$^{1}$,  Ross J.\ McLure$^{1}$, 
\newauthor Duncan Farrah$^{6}$
\\  
    $^{1}$SUPA, Institute for Astronomy, Royal Observatory Edinburgh,
    Blackford Hill, Edinburgh, EH9~3HJ, UK\\
    $^{2}$Institute for Computational Cosmology, Durham University,
    South Road, Durham, DH1~3LE, UK\\
    $^{3}$Astronomy Technology Centre, Royal Observatory Edinburgh,
    Blackford Hill, Edinburgh, EH9 3HJ, UK\\
    $^{4}$Astrophysics, Department of Physics, Keble Road, Oxford,
    OX1~3RH, UK\\
    $^{5}$Space Science and Technology, Rutherford Appleton
    Laboratory, HSIC, Didcot, OX11~0QX, UK\\
    $^{6}$Astronomy Centre, Department of Physics and Astronomy,
    University of Sussex, Brighton, BN1~9QJ, UK
}
\date{\today}
\pagerange{\pageref{firstpage}--\pageref{lastpage}; } \pubyear{2009}
\label{firstpage}
\maketitle

\begin{abstract}
We compare H$\alpha$, ultraviolet (UV) and infrared (IR) indicators of
star formation rate (SFR) for a well-defined sample of $z=0.84$
emission-line galaxies from the High-$z$ Emission Line Survey
(HiZELS).  Using emission-line, optical, IR, radio and X-ray
diagnostics, we estimate that 5 -- 11~per~cent of H$\alpha$ emitters
at this redshift are active galactic nuclei.  We detect 35~per~cent of
the H$\alpha$ emitters individually at 24~$\mu$m, and stack the
locations of star-forming emitters on deep 24-$\mu$m {\it Spitzer
Space Telescope} images in order to calculate the typical SFRs of our
H$\alpha$-emitting galaxies.  These are compared to the observed
H$\alpha$ line fluxes in order to estimate the extinction at $z=0.84$,
and we find a significant increase in dust extinction for galaxies
with higher SFRs.  We demonstrate that the relationship between SFR
and extinction found in the local Universe is also suitable for our
high-redshift galaxies, and attribute the overall increase in the
typical dust extinction for $z=0.84$ galaxies to an increase in the
average SFR, rather than to a change in dust properties at higher
redshift.  We calculate the UV extinction, and find a similar
dependence on SFR to the H$\alpha$ results, but no evidence for a
2175~\AA\ UV bump in the dust attenuation law for high-redshift
star-forming galaxies.  By comparing H$\alpha$ and UV indicators, we
calculate the conversion between the dust attenuation of nebular and
stellar radiation, $\gamma$, and show that $\gamma=0.50\pm0.14$.  The
extinction / SFR relationship is shown to be applicable to galaxies
with a range of morphologies and bulge-to-disk ratios, to both merging
and non-merging galaxies, and to galaxies within high- and low-density
environments, implying that it is a fundamental property of
star-forming regions.  In order to allow future studies to easily
correct for a SFR-dependent amount of dust extinction, we present an
equation to predict the extinction of a galaxy, based solely upon its
observed H$\alpha$ luminosity, and use this to recalculate the
H$\alpha$ luminosity function and star formation rate density at
$z=0.84$.
\end{abstract}


\section{Introduction}
\label{sec:introduction}
There are many methods of estimating the global star formation rate
(SFR) of a galaxy, from measurements of the ultra-violet (UV) emission
coming directly from visible young massive stars, to the bolometric
infrared (IR) luminosity resulting from dust-reprocessed stellar
emission \citep[see][for a review of various star formation
indicators]{Kennicutt98}.  All SFR indicators suffer from different
biases and uncertainties, and while SFRs estimated in different ways
in the local Universe are now in general agreement, there is much
greater scatter in the results at $z\sim1$ \citep[see,
e.g.][]{Hopkins06}.

Locally one of the best measures of the current rate of star formation
within a galaxy comes from observations of the H$\alpha$ transition of
atomic hydrogen.  The luminosity of this recombination line scales
directly with the rate of ionising flux from young, massive stars and
gives an excellent indication of the presence of ongoing star
formation.  However, its usefulness as a quantitative SFR indicator is
reduced by the effect of dust obscuration; observed H$\alpha$ line
fluxes need to be corrected for extinction before the intrinsic
H$\alpha$ luminosity can be recovered, and the SFR calculated.

The initial method used to estimate H$\alpha$ extinction came from a
comparison of thermal radio emission, which is unaffected by dust
attenuation, and the observed H$\alpha$ + [N\,{\sc ii}] line emission
within local galaxies \citep{Kennicutt83}.  Disentangling thermal and
non-thermal radio emission is non-trivial \citep[see, e.g.][for more
details]{Condon92}, as thermal radio emission is only a small fraction
of the total radio luminosity of a galaxy \citep[$\lesssim10$~per~cent
at 1.4~GHz;][]{Condon90}, and so this method is not often used in
practise.

A much more common method to estimate extinction in individual
galaxies comes from the `Balmer decrement', comparing the observed
value of the H$\alpha$/H$\beta$ line flux ratio to the intrinsic value
when no dust is present.  This allows a calculation of the reddening
between the wavelengths of the H$\alpha$ and H$\beta$ lines and, with
a choice of dust attenuation model \citep[e.g.][]{Calzetti00}, permits
an estimate of the attenuation at a given wavelength \citep*[see
e.g.][]{Moustakas06}.  However, this technique requires sensitive
spectroscopy to obtain the line flux ratio with a good signal-to-noise
ratio (SNR), which is time-consuming for large samples of galaxies at
moderate redshifts.  It is therefore more usual for studies with
limited available photometry to simply assume that a single extinction
can be applied to all galaxies.  The canonical amount of H$\alpha$
extinction for a typical galaxy is usually taken to be $A_{\rm
H\alpha} \sim1$~mag \citep[e.g.][]{Kennicutt83,Bell01,
Pascual01,Charlot02,Fujita03,Brinchmann04,Ly07,Geach08,Sobral09Halpha},
based upon measurements of extinction in the local Universe.

At higher redshift, the appropriate H$\alpha$ extinction for a typical
galaxy is not well determined.  Significant numbers of extremely dusty
galaxies have been discovered at high redshift
\citep[e.g.][]{Chapman05}, and the large dust content implies a
greater H$\alpha$ attenuation.  The global rate of star formation
increases out to $z\sim1$ \citep[e.g.][]{Hopkins06}, and galaxies with
higher SFR are thought to undergo greater extinction \citep{Hopkins01,
Sullivan01, Berta03}, also implying that the typical H$\alpha$
attenuation will be greater at $z\sim1$ than in the local Universe.
\citet{Villar08} have estimated that the typical H$\alpha$ extinction
of galaxies within their $z=0.84$ sample is $\sim1.5$~mag.

The total IR luminosity of a galaxy is a bolometric measurement of the
thermal emission from dust grains, which have been heated by the UV
and optical radiation produced by young massive stars
\citep[e.g.][]{Kennicutt98}.  For star-forming galaxies with a
reasonable dust content, this emission gives a direct measurement of
the total stellar power output.  However, since longer wavelength IR
emission will more closely trace the cool dust component within a
galaxy, which predominantly results from heating by an old stellar
component rather than from recent star formation, it could be argued
that mid-IR emission might be a better SFR indicator than the
bolometric IR luminosity \citep[e.g.][]{Dopita02}.

\citet{Rieke09} have demonstrated that the flux density measured in
the {\it Spitzer Space Telescope} 24-$\mu$m band can be used as a good
proxy for SFR for galaxies at $z=0$, and, after applying suitable
corrections, has a comparable accuracy to estimating SFRs from the
extinction-corrected Pa$\alpha$ luminosity or total IR luminosity for
normal galaxies in the redshift range $0<z<2$.  Observations at
24~$\mu$m have the additional advantage over longer wavelength IR
observations in that they are more sensitive (thus allowing fainter,
more distant galaxies to be studied), and have a higher resolution
($\sim6$~arcsec, making it relatively straightforward to identify
counterparts at other wavelengths; the 70- and 160-$\mu$m {\it
Spitzer} bands only have resolutions of $\sim20$ and $\sim40$~arcsec
respectively).  These disadvantages will be overcome to some extent
with {\it Herschel}, which should be able to make sensitive
$\sim6$~arcsec resolution images of galaxies at 70~$\mu$m, close to
the peak in luminosity for the spectral energy distribution (SED) of a
typical galaxy at $z\sim0$.

In this work, we compare H$\alpha$, UV and IR SFR indicators in order
to estimate the extinction for a well-defined sample of galaxies at
$z=0.84$, using data from the High-$z$ Emission Line Survey
\citep[HiZELS;][]{Geach08, Sobral09Halpha, Sobral09Lyalpha}.  HiZELS
is a panoramic extragalactic survey making use of the wide area
coverage of the Wide Field Camera on the 3.8-m UK Infrared Telescope
(UKIRT).  Narrow-band observations of the Cosmological Evolution
Survey field \citep[COSMOS;][]{Scoville07overview}, and the
Subaru-XMM-UKIDSS Ultra Deep Survey field \citep[UDS;][]{Lawrence07}
have been taken in the $J$, $H$ and $K$ bands in order to target
H$\alpha$ emitters at $z=0.84$, $z = 1.47$ and $z=2.23$ respectively;
this paper concentrates on the properties of the $z=0.84$ H$\alpha$
sample, the creation of which is described in full in
\citet{Sobral09Halpha}.

In Section~\ref{sec:sampleselection} we present details of the HiZELS
sample selection, and the available multi-wavelength data in the
COSMOS and UDS fields.  In Section~\ref{sec:agn} we describe various
methods which we use to identify active galactic nuclei (AGN)
contaminants, and quantify the fraction of H$\alpha$-emitting sources
which may be AGN at $z=0.84$.  We describe our method for stacking
galaxies and measuring typical 24-$\mu$m flux densities in
Section~\ref{sec:results}, and convert our stacking results into SFRs
in order to estimate the H$\alpha$ extinction that is required to make
this indicator agree with the IR-based SFRs.  In
Section~\ref{sec:discussion} we discuss our results, investigate the
effects that a variable extinction has on the H$\alpha$ luminosity
function at $z=0.84$, and present a method for correcting observed
H$\alpha$ luminosities for a SFR-dependent amount of extinction.
Section~\ref{sec:conclusions} reports our conclusions.

Throughout this work we assume a concordance cosmology of $\Omega_{\rm
M}=0.3$, $\Omega_{\Lambda}=0.7$ and H$_{0}=70$~km~s$^{-1}$~Mpc$^{-1}$.
All magnitudes are given in the AB system \citep{Oke83}.

\section{HiZELS sample selection}
\label{sec:sampleselection}
We give a brief overview of the HiZELS sample selection procedure
below; full details can be found in \citet{Sobral09Halpha}.
Emission-line candidates were selected on the basis of a flux excess
in a narrow $J$-band filter centred on $\lambda=1.211$~$\mu$m,
compared with deep $J$-band continuum observations.  Spurious sources
(mainly artifacts due to bright stars) were cleaned from the sample,
and spectroscopic and photometric redshifts were used to identify and
remove sources found to be a different line emitter at another
redshift (such as H$\beta$ or [O\,{\sc iii}] at $z\sim1.5$).

The H$\alpha$ line flux for each source was calculated from the
measured flux densities in the narrow and broad-band filters, as
described in \citet{Sobral09Halpha}.  No aperture corrections were
applied to the H$\alpha$ flux, since the correction factor is expected
to be minimal.  Contamination from the nearby [N\,{\sc ii}] lines at
6548 and 6583~\AA\ was corrected for, using the relationship between
the flux ratio $F_{[{\rm N\,II}]}/F_{\rm H\alpha}$ and the total
measured equivalent width EW(H$\alpha$+[N\,{\sc ii}]) from
\citet{Villar08}, with the median correction factor being 25~per~cent.

The final HiZELS H$\alpha$ sample consisted of 477 H$\alpha$-emitting
sources in COSMOS and 266 in UDS, over a total area of 1.30~deg$^{2}$,
at a redshift of $z=0.845\pm0.021$, down to an average observed
$3\sigma$ H$\alpha$ line flux limit of
$8\times10^{-17}$~erg~s$^{-1}$~cm$^{-2}$.

\subsection{The HiZELS COSMOS sample}
Ground-based observations of the COSMOS field have been made at a
variety of UV, optical and near-IR wavelengths, and \citet{Capak07}
have produced a point-spread function (PSF)-matched catalogue for all
sources with $I$-band magnitude $<25$.  This catalogue contains data
from observations with the Canada France Hawaii Telescope (CFHT;
$u^{*}$ and $i^{*}$ bands\footnote{Throughout this work we retain the
notation of \citet{Capak07} to differentiate between observations in
different filter systems.}), the Sloan Digital Sky Survey \citep[SDSS
$ugriz$ bands;][]{Abazajian04} and the Subaru telescope ($B_{J}$,
$V_{J}$, $g^{+}$, $r^{+}$, $i^{+}$, $z^{+}$ and NB816 bands), along
with $K_{\rm s}$-band data taken from a combination of observations
from the Cerro Tololo International Observatory and the Kitt Peak
National Observatory.  Further details on the catalogue construction
can be found in \citet{Capak07}.  All of the COSMOS H$\alpha$ emitters
have counterparts in this PSF-matched catalogue.

Mid-IR and far-IR observations of COSMOS have been taken as part of
the {\it Spitzer}-COSMOS survey
\citep[S-COSMOS;][]{Sanders07,LeFloch09}.  464 of the H$\alpha$
emitters (97~per~cent) have a counterpart in Infrared Array Camera
(IRAC) observations of the field, in at least one of the 3.6-, 4.5-,
5.8- and 8-$\mu$m bands.  Several of our AGN classification
diagnostics rely upon the availability of IRAC information (see
Section~\ref{sec:agn}); for this reason, we reject the 13 sources
without IRAC detections from our study.  To determine whether there is
any potential bias from this exclusion, we repeated all tests in
Section~\ref{sec:results} while retaining these sources -- no
significant differences were seen in our results.

There have been several {\it Spitzer} 24-$\mu$m surveys of the COSMOS
field, and we have used the most recent of these, the Multiband
Imaging Photometer for {\it Spitzer} (MIPS) Cycle~3
observations\footnote{available via
http://www.ifa.hawaii.edu/$\sim$ilbert/S-Cosmos/}, in this work.  The
public source catalogue is cut at a flux density limit of 0.15~mJy
($\sim7\sigma$), and 158 H$\alpha$ emitters had a counterpart in this
catalogue, where we consider that an H$\alpha$ emitter has a
counterpart if the positions of the 24-$\mu$m and optical source
centres are within 3~arcsec.  We estimate the percentage of H$\alpha$
emitters which may have an incorrect 24-$\mu$m association by shifting
their positions by $+10$~arcmin in declination and recalculating the
number of matches -- nine spurious associations were made, which
implies an overall percentage of H$\alpha$ emitters with individual
24-$\mu$m counterparts of $34\pm2$~per~cent.  Throughout this work,
all matching percentages have been corrected for false matches in this
way.

1.4-GHz radio observations of the COSMOS field have been taken with
the Very Large Array \citep[VLA;][]{Schinnerer04, Schinnerer07,
  Bondi08}.  The H$\alpha$ emitters are located within the VLA-COSMOS
Deep Project area, which has a typical noise level of between 10 and
40~$\mu$Jy~beam$^{-1}$, and 21 H$\alpha$ emitters
($4.5\pm0.9$~per~cent) have 1.4-GHz counterparts within 3~arcsec.  The
full HiZELS region of the COSMOS field has been observed with {\it
  XMM-Newton} \citep{Cappelluti09}, and the majority of the region has
been observed with {\it Chandra} \citep{Elvis09} -- 7
($1.5\pm0.4$~per~cent) and 8 ($1.7\pm0.2$~per~cent) H$\alpha$ emitters
have X-ray counterparts within 3~arcsec respectively, with 5 H$\alpha$
emitters having counterparts from both surveys.

\subsection{The HiZELS UDS sample}
\label{sec:udssample}
The UDS field is the location of a very deep UKIRT near-IR survey in
the $J$, $H$ and $K$~bands \citep{Warren07DR1}.  There is additional
deep imaging data available from Subaru \citep[$B_{J}$, $V_{J}$,
$r^{+}$, $i^{+}$, $z^{+}$ bands;][]{Furusawa08}, and $u^{*}$-band data
from CFHT.

A deep {\it Spitzer} legacy survey of the UDS field has been carried
out with both IRAC and MIPS (SpUDS; P.I.\ J.\ Dunlop).  Catalogue
production is still ongoing, but mosaics of the SpUDS survey have been
delivered to the community\footnote{available via
http://ssc.spitzer.caltech.edu/legacy/spudshistory.html}.  We use a
preliminary version of the IRAC catalogue (M.\ Cirasuolo et al., in
preparation), which excludes regions of the UDS image near to bright
stars, image artifacts or the mosaic edges -- 234 of the H$\alpha$
emitters are in regions covered by this catalogue.  We exclude a
further four sources from our analysis because they are either outside
or on the edge of the SpUDS 24-$\mu$m image, leaving 230 sources, all
of which had optical / IR counterparts.  Photometry for all bands was
measured within a 1.8~arcsec diameter aperture, equivalent to roughly
$2\times$ the size of point spread function for the optical and
near-IR images.

We make use of the finely sampled 24-$\mu$m SpUDS image of the field
(1.245~arcsec/pixel).  This image has asteroid streaks visible; we
inspected the locations of the H$\alpha$ emitters to confirm that
these imaging artifacts would not affect our analysis.  84 of the UDS
emitters ($37\pm3$~per~cent) had counterparts in the preliminary
24-$\mu$m source catalogue, which we have cut at 0.1~mJy
($\sim6\sigma$), using the same matching criteria as for the COSMOS
field.

Radio observations of the UDS field have been taken at 1.4~GHz
\citep{Simpson06}, and three of the H$\alpha$ emitters
($1.3\pm0.4$~per~cent) have 1.4-GHz counterparts above 100~$\mu$Jy.
{\it XMM-Newton} observations of the field have been taken as part of
the larger Subaru/{\it XMM} Deep Survey \citep{Ueda08}, and four of
the H$\alpha$ emitters ($1.7\pm0.4$~per~cent) have X-ray counterparts.

\section{Identification of AGN contaminants}
\label{sec:agn}
The HiZELS selection process identifies H$\alpha$ emitters within the
redshift range of $z=0.845\pm0.021$.  While most of these sources are
likely to be powered by star formation, there will be some objects
where AGN activity dominates the energetics -- SFRs estimated from
these sources will not be accurate, and these contaminants need
removing from the sample.  We adopt a number of methods for
classifying sources as `possible' or `likely' AGN, with the number of
sources satisfying each of the criteria listed in
Table~\ref{tab:agnsummary}.

\subsection{SED template fitting}
\label{sec:sed}
\begin{figure}
  \begin{center}
    \includegraphics[width=0.45\textwidth]{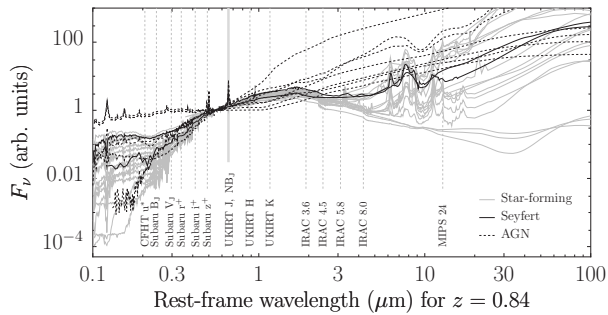}
    \caption{The template SEDs used to fit the H$\alpha$ emitters,
    taken from \citet{Dale01} and \citet{Polletta07} and shown in the
    galaxy rest-frame, arbitrarily normalised to a flux density of 1
    at 5500~\AA.  The vertical dashed lines indicate the effective
    wavelengths of the available photometry for sources at $z=0.84$,
    with the narrow/broad-band $J$ selection shown by the thicker grey
    line.  Note the difference in AGN and star-forming galaxy mid-IR
    SED shapes.}
    \label{fig:sedshapes}
  \end{center}
\end{figure}
The matched optical and IR photometry available for each H$\alpha$
emitter was described in Section~\ref{sec:sampleselection}.  This
photometry was blue-shifted to the source rest-frame and fitted to a
series of template SEDs \citep[shown in
Fig.~\ref{fig:sedshapes};][]{Dale01,Polletta07}, using a standard
$\chi^{2}$ minimization procedure.  Only the data with the highest SNR
was used out of multiple observations at similar wavelengths, such as
the SDSS $u$ and CFHT $u^{*}$~bands.  The SEDs consisted of
star-forming galaxies (spiral, elliptical and starburst templates),
Seyferts, and AGN templates (Type-1, Type-2, and two composite
starburst/AGN templates).  Sources which were best fitted by an AGN
SED were classified as `possible' AGN candidates -- we do not classify
sources as `likely' AGN through this automatic fitting procedure, as
the range of SED templates is limited to a few per source type, and
may not fully represent the HiZELS population.
Fig.~\ref{fig:sedshapes} shows that while there is a large amount of
photometric information available for the H$\alpha$ emitters, there is
very little difference between the colours of the Seyfert and
Starburst templates where photometry is available -- for this reason,
sources that were best-fitted by Seyfert templates were excluded from
this automatic classification.

The SEDs of all sources were inspected to confirm the results of the
automatic fitting, and to identify any further objects which appeared
to be potential AGN, but had not been classified as such.  Four
further sources were manually classified as possible AGN; these
typically had a similar shape to the AGN templates but had an offset
in the IR to optical flux densities, and for this reason were not
automatically classified as AGN.

\subsection{Colour-colour diagnostics}
\label{sec:agnirselection}
We use a number of colour-colour diagnostics to identify further AGN
contaminants.  For each, we classified sources located within the
selection region as `likely' AGN, while those located within the
region, but with errors on their flux densities sufficiently large
enough to potentially place them outside the region, were classified
as `possible' AGN.  

\begin{figure*}
  \begin{center}
    \includegraphics[width=0.9\textwidth]{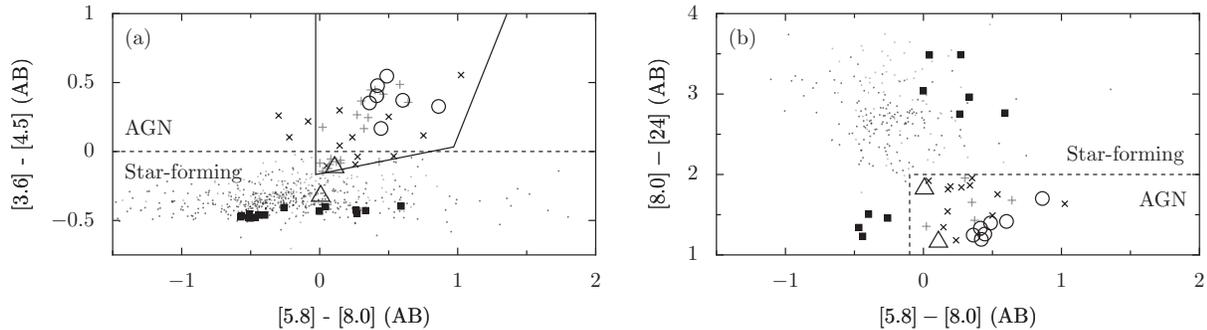}
    \caption{Infrared colour-colour plots showing the HiZELS H$\alpha$
    emitters (COSMOS galaxies with black dots, UDS galaxies with grey
    dots; AGN candidates with black diagonal crosses and grey upright
    crosses respectively).  AGN selection regions are indicated by the
    lines: (a) the \citet{Stern05} criteria is shown with the solid
    line, and the additional 3.6- to 4.5-$\mu$m criterion is indicated
    by the dashed line; (b) the region $R(5.8, 8, 24)$ is indicated by
    the dashed line.  The colours of the SEDs shown on
    Fig.~\ref{fig:sedshapes} are also plotted for reference, separated
    into star-forming galaxies (solid black squares), Seyferts (open
    triangles) and AGN (open circles).}
    \label{fig:agnircolours}
  \end{center}
\end{figure*}

We used the mid-IR selection criteria of \citet{Stern05} to define a
region of IRAC colour-colour space that principally contains AGN,
which is shown on Fig.~\ref{fig:agnircolours}(a), along with the
positions of the H$\alpha$ emitters with detections in all four IRAC
bands.  The objects which have been flagged as potential AGN
contaminants lie well away from the remainder of the population.

\citet{Stern05} warn that their criteria may preferentially omit AGN
at $z\simeq0.8$, where the difference in mid-IR colours between
star-forming galaxies and AGN becomes less distinct.
Fig.~\ref{fig:sedshapes} shows that the composite black-body spectrum
of galaxies powered by star formation peaks at around 1.6~$\mu$m, and
falls at longer wavelengths.  Warm dust located near to an AGN
increases the luminosity of the mid-IR continuum, and
Fig.~\ref{fig:sedshapes} indicates that the AGN templates continue to
rise with wavelength throughout the mid-IR.  For this reason we add a
further AGN classification for sources that are outside of the
\citet{Stern05} region, but have a rising SED between the 3.6- and
4.5-$\mu$m bands (which probe rest-frame 1.9- to 2.4-$\mu$m dust
emission), which is also shown on Fig.~\ref{fig:agnircolours}(a).

It is difficult to conclusively discriminate between Seyfert and
starburst galaxies, given the available photometry -- the optical SED
shape of the two source types can be very similar, with the main
difference being the increased amount of 24-$\mu$m emission for
starbursts.  We use the shape of the mid-IR to far-IR SED to define a
region $R(5.8, 8, 24)$ that contains Seyferts, from inspection of
Fig.~\ref{fig:sedshapes};
\begin{equation}
R(5.8, 8, 24) \equiv ([8] - [24] < 2~{\rm and}~ [5.8] - [8] > -0.1).
\end{equation}
The first criterion distinguishes between Seyferts and starbursts, and
the second criterion prevents more quiescent source types such as
spirals and elliptical galaxies from being flagged.
Fig.~\ref{fig:agnircolours}(b) shows the selection region for these
criteria, which also includes the other AGN SEDs shown on
Fig.~\ref{fig:sedshapes}.

\begin{table}
  \begin{center}
  \caption{The number of H$\alpha$ emitters classified as `possible'
  or `likely' AGN under each of the criteria described in
  Section~\ref{sec:agn}.  Several sources fulfilled more than one
  criterion, so the total number of sources classified as AGN is less
  than the sum of the individual classifications.  Sources with two or
  more `possible' classifications were considered to be `likely' AGN.
  Values in parentheses represent the number of sources which were
  only flagged as AGN under a single criterion.}
  \label{tab:agnsummary}
    \begin{tabular}{lcccc}
\hline
Method & Type & COSMOS & UDS & Total\\
\hline
Automatic SED fit              & Possible & 17 (10) & 20 (10) & 37 (20)\\
Manual SED fit                 & Possible &  4 (4) &  0 (0) &  4 (4)\\
Stern wedge                    & Possible &  3 (2) & 17 (8) & 20 (10)\\
Stern wedge                    & Likely   &  7 (0) &  2 (0) &  9 (0)\\
$[3.6] - [4.5]$ colour         & Possible &  1 (1) &  2 (1) &  3 (2)\\
$[3.6] - [4.5]$ colour         & Likely   &  4 (3) &  0 (0) &  4 (3)\\
Mid- to far-IR slope           & Possible &  3 (3) &  2 (0) &  5 (3)\\
Mid- to far-IR slope           & Likely   &  9 (1) &  5 (0) & 14 (1)\\
X-ray counterpart              & Likely   & 10 (3) &  4 (0) & 14 (3)\\
Low value of $q_{24}$          & Likely   &  1 (0) &  3 (3) &  4 (3)\\
Emission-line ratio      & Likely   &  $\geqslant3$ (2)$^{a}$ &  -
&  $\geqslant3$ (2)$^{a}$\\ 
\hline
H$\alpha$ emitters             & & 464 & 230 & 694\\
Possible AGN                   & &  20 &  19 &  39\\
Likely AGN                     & &  22 &  15 &  37\\
AGN contamination              & & 5 -- 9\% & 7 -- 15\% & 5 -- 11\% \\
\hline
\multicolumn{2}{l}{Sources retained for analysis} & 422 & 196 & 618\\
\hline
    \end{tabular}
  \end{center}
$^{a}$ This value is a lower limit, as only 26 COSMOS sources
  (6~per~cent) had spectroscopy with sufficient SNR for this method.
\end{table}

\subsection{X-ray detections}
The COSMOS field has been observed with both {\it XMM-Newton} and {\it
Chandra} \citep{Ueda08,Elvis09}, and the UDS field has been observed
with {\it XMM-Newton} \citep{Cappelluti09}.  The catalogue flux limits
are equivalent to luminosity limits in the 2--10~keV energy band of
$3\times10^{43}$~erg~s$^{-1}$, $2.5\times10^{42}$~erg~s$^{-1}$ and
$1\times10^{43}$~erg~s$^{-1}$ respectively, for sources at $z=0.845$.
Purely star-forming galaxies can emit X-rays, but locally are not
observed to have luminosities exceeding $L_{0.5-8~{\rm keV}} \approx
3\times10^{42}$~erg~s$^{-1}$ \citep{Bauer04} -- all of the H$\alpha$
emitters with an X-ray detection have luminosities significantly above
this value, and are classified as `likely' AGN.

\subsection{Radio-loud sources}
We use the IR / radio correlation \citep[e.g.][]{Appleton04} to
distinguish between radio emission resulting from AGN and
star-formation activity, for the H$\alpha$ emitters with radio
counterparts.  22/24 of these had catalogued 24-$\mu$m counterparts --
visual inspection of the MIPS images confirmed the lack of a
counterpart for the remaining two sources, and we place a conservative
upper limit of 0.1~mJy on the 24-$\mu$m flux density of these sources.

The logarithmic ratio of 24-$\mu$m and 610-MHz flux densities is
$0.55\pm0.4$ for star-forming galaxies at $z\sim0.8$ which are
detected at both wavelengths \citep[see][for further
details]{Garn09SFR}; converting this to a 1.4-GHz value using
equation~(4) of \citet{Garn09stacking}, and assuming a radio spectral
index of 0.8, we obtain a typical value of $q_{24} \equiv {\rm
  log}_{10}(S_{24}/S_{1.4}) = 0.8\pm0.4$ for
star-forming galaxies at $z=0.84$.  We find that indeed most of the
HiZELS radio emitters do cluster around $q_{24}\sim0.7$, but four lie
well away from this, with $q_{24}<0.1$, and these are classified as
`likely' AGN.

\subsection{Emission-line ratios}
\label{sec:emissionline}
Spectroscopic data is available for 20~per~cent of the COSMOS H$\alpha$
emitters from the $z$-COSMOS survey \citep{Lilly09}, and
\citet{Sobral09Halpha} used the [O\,{\sc ii}]3727/H$\beta$ and
[O\,{\sc iii}]5007/H$\beta$ emission-line ratios to separate AGN from
star-forming galaxies.  26 of the emitters had spectroscopy at
sufficient SNR to carry out this comparison; of these sources, three
(12~per~cent) were classified by \citet{Sobral09Halpha} as likely AGN
contaminants, and we retain that classification in this work.  No
spectroscopy is currently available for the UDS HiZELS sample, and as
we do not have complete spectroscopy for the entire COSMOS sample,
this value is a lower limit.

\subsection{The AGN contamination rate at $z = 0.84$}
\label{sec:agnsummary}
Several of the H$\alpha$ emitters were classified as AGN under more
than one criteria, and sources with two or more `possible'
classifications were considered to be `likely' AGN candidates.  In
practice the distinction between `possible' and `likely' AGN
candidates is only of interest in quantifying the uncertainty in the
fraction of H$\alpha$ emitters which may be AGN; we do not use either
of these types of source in later analysis.  The total `possible' and
`likely' AGN in each field were 20 and 22 (COSMOS) and 19 and 15
(UDS).  We find a significantly different AGN fraction between the two
fields (5 -- 9~per~cent in COSMOS, and 7 -- 15~per~cent in UDS).  The
principal cause of this discrepancy is the number of sources
classified as AGN through the IR colour-colour criteria; 10/464
(2.1~per~cent) of COSMOS sources and 19/230 (8.3~per~cent) of UDS
sources fell within the \citet{Stern05} region.  We have placed more
conservative errors on the UDS photometry as production of a
band-merged catalogue is still ongoing -- this leads to a higher
fraction of UDS sources being `possible' AGN than in COSMOS.

We visually inspected the SEDs of these sources in order to
investigate this discrepancy -- the majority showed a clear power-law
AGN shape, and were also flagged as AGN under at least one other
criteria, making it very likely that the classifications are correct.
We believe that the bulk of the discrepancy between the two fields may
be due to cosmic variance, although the slightly greater errors on the
UDS photometry (which may incorrectly move a few sources into the AGN
classification regions) may exaggerate this.

By combining the samples, we obtain an overall AGN contamination rate
of between 5 and 11~per~cent.  The majority of sources classified as
`possible' AGN contaminants were best-fitted by an AGN template, but
did not satisfy any of the specific colour requirements; for this
reason, we believe that the true AGN contamination rate will be near
to the upper end of our quoted range.  The variation in the number of
sources detected by each technique suggests that (i) using a single
criterion for starburst / AGN discrimination will underestimate the
true AGN fraction, and (ii) a combination of colour-colour methods and
full SED fitting should be used in order to maximise the success rate
for AGN identification.

The technique of source stacking (used in Section~\ref{sec:results})
implicitly assumes that the sources being combined statistically are
comparable to each other, but it is robust to the presence of a
small number of outlier sources, such as AGN contaminants
\citep[see][for further details]{Garn09stacking}.  The results in the
remainder of this work use only those sources that were not classified
as either likely or possible AGN, but we have repeated all tests while
retaining the possible AGN in our sample, and found no significant
difference in any of the results.  The H$\alpha$ and 24-$\mu$m flux
density of the AGN contaminants are very weakly correlated (see
Section~\ref{sec:resultsflux}), and we believe that it is unlikely that
any remaining AGN contaminants should have a significant effect on the
results presented in this work.

\begin{figure*}
  \begin{center}
    \includegraphics[width=0.9\textwidth]{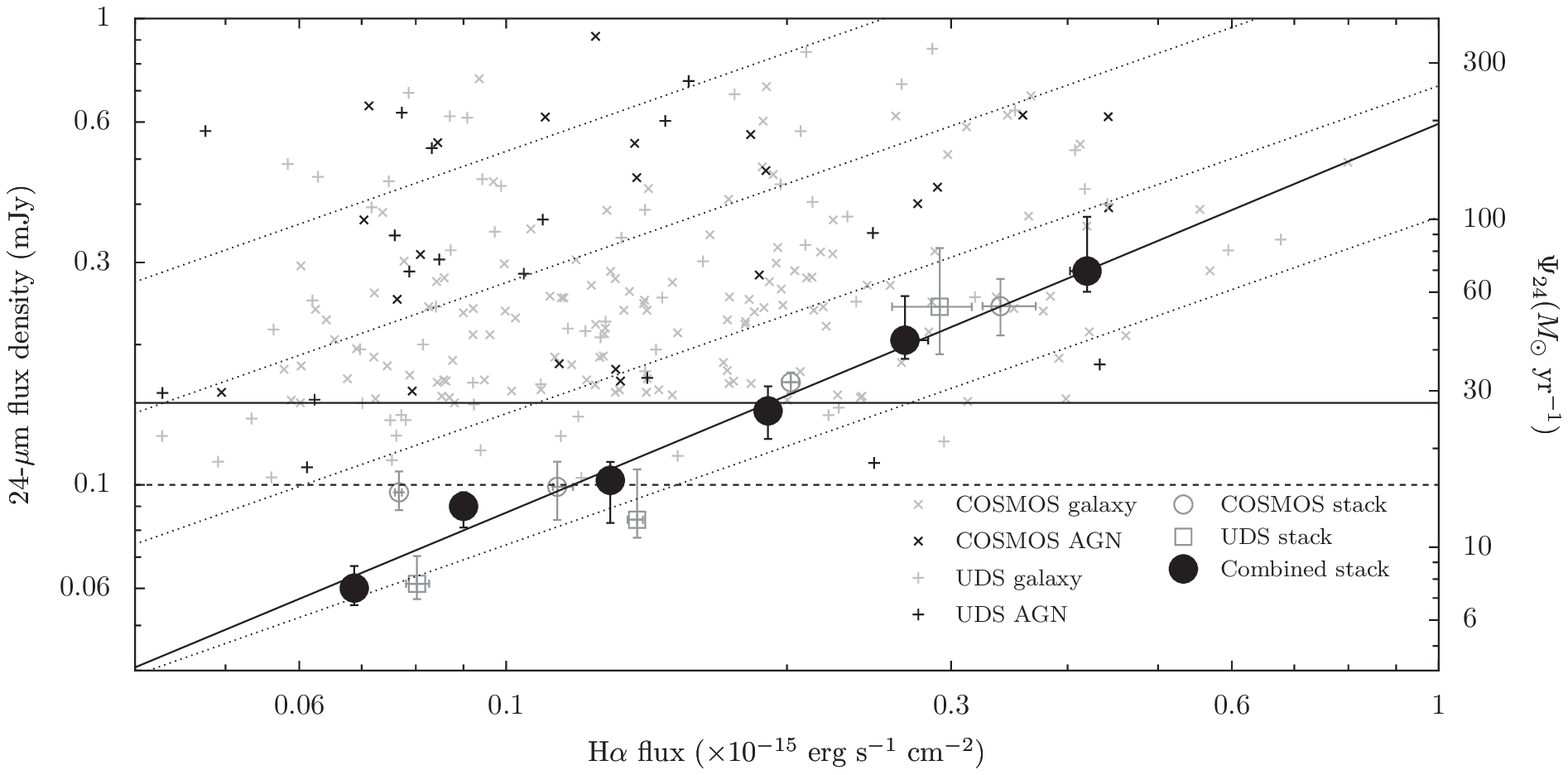}
    \caption{Variation in 24-$\mu$m flux density with H$\alpha$ flux,
    for the stacks of all H$\alpha$ emitters which were not flagged as
    AGN.  The solid diagonal line represents the best-fitting
    relationship between 24-$\mu$m flux density and H$\alpha$ flux,
    with a power-law slope of 0.83.  Individual H$\alpha$ emitters
    with 24-$\mu$m detections are indicated for comparison with the
    stacked data, separated by their AGN classification, and the
    COSMOS catalogue limit of 0.15~mJy ($\sim7\sigma$) and UDS flux
    limit of 0.1~mJy are indicated by the horizontal solid and dashed
    lines respectively.  Note that the filter profile correction,
    described in Section~\ref{sec:resultsflux}, has been applied to
    the stacked H$\alpha$ fluxes but not to the individual data
    points.  The right-hand axis indicates the corresponding SFR for
    each 24-$\mu$m flux density, calculated using
    equation~(\ref{eq:irsfr}).  The diagonal dotted lines represent
    the \citet{Kennicutt98} relationship between H$\alpha$ luminosity
    and SFR, modified by a factor of 0.66 in order to make the IMFs
    agree (see Section~\ref{sec:sfrcalculation}), and assuming an
    extinction of 1, 2, 3 and 4~mag (bottom to top). }
    \label{fig:halphafluxflux}
  \end{center}
\end{figure*}
\section{Analysis \& Results}
\label{sec:results}
\subsection{Calculation of stacked 24-$\mu$m flux densities}
\label{sec:resultsflux}
It is necessary to know the locations of sources to within the size of
a pixel (1.2 and 1.245~arcsec for the COSMOS and UDS 24-$\mu$m images
respectively) in order to use source stacking.  We have used source
positions taken from the optical catalogues, which are expected to
have an accuracy of about 0.2~arcsec \citep{Capak07,Furusawa08}, and
we verified that there is no significant coordinate offset between the
HiZELS data and their counterparts in the 24-$\mu$m images (median
offsets in right ascension and declination of $-0.04\pm0.04$~arcsec
and $0.03\pm0.04$~arcsec for COSMOS sources, and $-0.05\pm0.08$~arcsec
and $0.30\pm0.09$~arcsec for UDS sources).  The small positional
offsets were not corrected for in later analysis.

The stacking technique used is very similar to that described in
\citet{Garn09stacking}, and further details and validation of the
procedure can be found in that work.  Sources were binned by their
observed H$\alpha$ line flux, and a median H$\alpha$ flux calculated
for each bin.  A typical 24-$\mu$m flux density was then calculated
for each set of binned data in two different ways:

\begin{enumerate}
\item The 24-$\mu$m flux density of each H$\alpha$ emitter was
  measured within a circular aperture of 3~arcsec radius, centred on
  the source location.  While individual flux density measurements are
  affected by the local 24-$\mu$m noise level, the statistical
  properties of the distribution give a robust estimate of the typical
  flux density for sources within the bin.  The median was used as the
  statistical estimator, to avoid being biased by the combination of
  detected and undetected sources \citep[see][for further
  details]{Garn09stacking}, as for many stacks, the median flux
  density will come from individual detections at 24~$\mu$m (see
  Fig.~\ref{fig:halphafluxflux}).

\item A cut-out 24-$\mu$m image was created for each source in a bin,
  centred on the known HiZELS locations.  From these cutouts, a
  stacked image was created, with each pixel calculated from the
  median of the values of the corresponding pixels of the cutout
  images.  The flux density of the stacked image was then measured in
  the same way as for method (i).  We experimented with the suggestion
  of \citet{Lonsdale09} of placing images into a stack twice, once at
  $90^{\circ}$ to the original orientation, in order to remove any
  systematic instrumental effects, but found that this made a
  negligible difference to the stacking results.
\end{enumerate}

Measurements of the flux density obtained from random positions
indicated the presence of a slightly negative background level in both
images (approximately $-0.01$~MJy~sr$^{-1}$, equating to a flux within
a 3~arcsec radius of $-8$~$\mu$Jy), which was subtracted from the flux
density measurements to avoid biasing the stacking results.

The measured 24-$\mu$m flux density was multiplied by an aperture
correction factor, obtained through comparison of the measured and
catalogued flux densities of 24-$\mu$m sources within the COSMOS
field, using a similar method to that in \citet{Garn09stacking}.  The
calculated correction factor was $3.41\pm0.12$, and should also be
appropriate for UDS -- the agreement between the stacked results from
the two fields (see Fig.~\ref{fig:halphafluxflux}) implies that we are
not introducing a systematic effect by using the COSMOS-derived
aperture correction for the UDS stacking.

The narrow $J$-band filter used to identify H$\alpha$ emitters is not
a perfect top-hat function, and this can have a measurable effect for
sources with H$\alpha$ emission at wavelengths near the edge of the
filter \citep[as discussed in][]{Sobral09Halpha}.  We calculate the
statistical effect of this by simulating a distribution of $10^{5}$
sources based upon the \citet{Sobral09Halpha} H$\alpha$ luminosity
function.  We spread them evenly in redshift over the range $z=0.81 -
0.87$, and use the actual filter profile to recover the luminosity
distribution.  Through comparison of the median input and median
recovered luminosity, a statistical correction to the observed
luminosity of each source was obtained -- these corrections were
applied to the H$\alpha$ data before stacking took place.  The filter
profile correction has a moderate effect on the faintest emitters (up
to a $\sim20$~per~cent increase in luminosity), and little effect for
the brightest emitters, as these can typically only be observed to be
bright if they are not detected in low transmission regions of the
filter.  Nevertheless, all of the trends discussed in this work remain
present if no correction is applied.

Error estimates for the H$\alpha$ fluxes and 24-$\mu$m stacked flux
densities indicate the 68 percent confidence limits ($\pm1\sigma$)
obtained from the measured flux distribution \citep{Gott01}.  The flux
densities obtained using both stacking methods were consistent to
within 3~per~cent, which we consider to be an additional uncertainty
introduced through the stacking process, and add in quadrature to our
error estimate.  Throughout this work, all stacking results have been
taken using method~(i).  The dominant errors on our stacked results
come from the uncertainty in estimating stacked flux densities, due to
the small number of galaxies in each bin (typically between $\sim20$
and $\sim160$).

\subsection{Variation in 24-$\mu$m flux density or SFR with observed H$\alpha$ flux}
\label{sec:sfrcalculation}
Fig.~\ref{fig:halphafluxflux} shows the stacked flux density
measurements using the combined data from the COSMOS and UDS fields,
with the results from each individual field also indicated for
reference.  While there is a strong correlation between the two
stacked flux measurements, the relationship is significantly
non-linear; the best-fitting power law has a slope of $0.83\pm0.05$,
and is indicated on Fig.~\ref{fig:halphafluxflux}.  We have plotted
individual H$\alpha$ emitters with 24-$\mu$m detections on
Fig.~\ref{fig:halphafluxflux}, separated by their AGN classification,
in order to investigate whether there is any strong dependence of
24-$\mu$m flux density on H$\alpha$ flux for the AGN sources with
individual 24-$\mu$m detections -- no such dependence is seen, with a
Pearson product-moment correlation coefficient of $\rho=0.29$,
compared with $\rho=0.55$ for the sources that are not classified as
AGN.

\citet{Rieke09} present a method for determining SFRs directly from
{\it Spitzer} 24-$\mu$m flux density measurements.  They assemble a
series of template SEDs for galaxies with different luminosities, and
create average SED templates which span the total IR luminosity range
of $5\times10^{9} - 10^{13}$~$L_{\odot}$.  From these bolometric
luminosities, a SFR is calculated using a similar relationship to that
in \citet{Kennicutt98}.  Each template is redshifted in steps of
$\Delta z = 0.2$, and convolved with the MIPS 24-$\mu$m instrumental
response function, in order to calculate the flux density which would
be observed for the galaxy at a given redshift.  The relationship
between {\it observed} 24-$\mu$m flux density and SFR is parameterised
at each discrete redshift, with the relationship at intermediate
redshifts available through interpolation of the fit coefficients.
The \citet{Rieke09} parameterisation takes into account the variation
in spectral shape (i.e.\ the K-correction) which is needed when
converting from an observation-frame 24-$\mu$m flux density into a
rest-frame 24-$\mu$m luminosity.  Accurate knowledge of the
K-correction is very important for 24-$\mu$m observations, which do
not just probe the hot dust emission from the galaxy, but can also
detect different polycyclic aromatic hydrocarbon features \citep*[PAH
molecules; see e.g.][]{Draine03} at varying redshifts.

At $z=0.845$, the \citet{Rieke09} relationship is
\begin{equation}
{\rm log}_{10}\left(\frac{\Psi_{24}}{M_{\odot}~{\rm yr}^{-1}}\right) =
6.8795 + 1.4224\times {\rm log}_{10}\left(\frac{S_{24}}{\rm
  Jy}\right),
\label{eq:irsfr}
\end{equation}
where $S_{24}$ is the flux density measured within the 24-$\mu$m MIPS
band, and we explicitly denote SFRs calculated from this band as
$\Psi_{24}$.  We have not included any systematic contribution to the
errors from uncertainties in this conversion between flux density and
SFR.

As with other methods of estimating SFR, there is an uncertainty due
to the conversion between an observable parameter that comes
principally from high-mass stars ($\gtrsim 5$~$M_{\odot}$, which carry
out the majority of the dust heating), and a SFR which includes the
rate of formation of much lower mass stars (down to 0.1~$M_{\odot}$).
The Initial Mass Function (IMF) used to convert between the two is
typically assumed to be a single power-law slope of $-1.35$ from 0.1
to 100~$M_{\odot}$ \citep{Salpeter55}, which can lead to an
over-estimate of the number of low-mass stars which are forming.
\citet{Rieke09} use a modified IMF with a shallower slope at low mass
\citep[$-0.3$ from 0.08 to 0.5~$M_{\odot}$ and $-1.3$ from 0.5 to
100~$M_{\odot}$;][]{Kroupa01} in order to calculate their SFR
estimates, which results in a SFR that is a factor of 0.66 times the
value which would be calculated from a Salpeter IMF containing the
same mass of high-mass stars.  This factor has already been applied to
the \citet{Rieke09} relationship in equation~(\ref{eq:irsfr}), and we
adopt it throughout the remainder of this work where appropriate -- in
particular, to modify the \citet{Kennicutt98} SFR relationships, which
were originally calculated using a Salpeter IMF.  Error estimates on
our results do not include any contribution from uncertainties in the
IMF.  However, all tracers of star formation that we consider in this
work (H$\alpha$, UV, IR) are most sensitive to high-mass stars, and a
change in assumed IMF at the low-mass end will make a very small
difference to the overall conclusions, as it would alter the
normalisation factors in the estimates of total SFR approximately
equally at all wavelengths.

We indicate the SFRs that correspond to a given 24-$\mu$m flux density
on Fig.~\ref{fig:halphafluxflux}, for reference.  The diagonal lines
indicate the \citet{Kennicutt98} relationship between H$\alpha$ flux
and SFR, assuming different levels of extinction for all galaxies; as
has been seen in previous works \citep[e.g.][]{Dopita02}, calculated
extinction values vary significantly from galaxy to galaxy.  The
stacked data has a typical extinction of between 1 and 2~mag for all
flux bins, and a visible increase for higher SFRs.  Note that the
individual galaxy detections are biased against low extinction (and
therefore low 24-$\mu$m SFR) sources since only H$\alpha$ emitters
with 24-$\mu$m counterparts are plotted -- the stacked data has no
such bias, as it uses both non-detections and sources individually
detected at 24~$\mu$m.

\subsection{Variation in H$\alpha$ extinction with observed H$\alpha$ luminosity}
\label{sec:halphaextinction}
\begin{figure*}
  \begin{center}
    \includegraphics[width=0.9\textwidth]{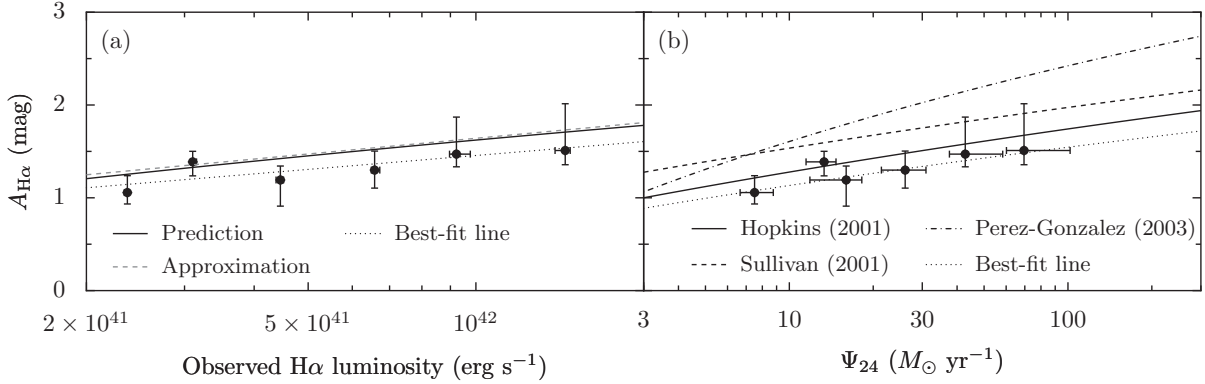}
    \caption{Variation in H$\alpha$ extinction for the H$\alpha$
    emitters (black circles with error bars), after binning them by
    their observed H$\alpha$ flux.  (a) Variation in extinction with
    observed H$\alpha$ luminosity (corrected for N~{\sc ii}
    contamination but not for dust attenuation).  The solid line shows
    the predicted relationship from equation~(\ref{eq:numerical}),
    with the approximation from equation~(\ref{eq:approximate}) shown
    by the dashed line.  The best fit to the data, from
    equation~(\ref{eq:datafit}), is indicated by the dotted line.  (b)
    Variation in extinction with 24-$\mu$m SFR.  The solid line shows
    the predicted relationship from
    equation~(\ref{eq:generalextinction}), using the fit between the
    Balmer decrement and SFR from \citet{Hopkins01}, while the dashed
    and dash-dot lines are the equivalent relationships, using the
    fits from \citet{Sullivan01} and \citet{PerezGonzalez03}
    respectively -- see Section~\ref{sec:halphaextinction} for further
    details.  The best fit to the data, using the slope of
    equation~(\ref{eq:generalextinction}) and $B=5.8$, is shown by the
    dotted line.  The hypothesis of a constant extinction for all
    H$\alpha$ luminosities or SFRs is excluded at the $3\sigma$ and
    $4\sigma$ level respectively. }
    \label{fig:halphasfrextinction}
  \end{center}
\end{figure*}
We can calculate the typical H$\alpha$ extinction for a bin of data,
if we are willing to assume that 24-$\mu$m emission for an
H$\alpha$-selected sample is a reasonable SFR indicator, as was found
by \citet{Rieke09}.  We invert the \citet{Kennicutt98} relationship
between SFR and H$\alpha$ luminosity (modified by a factor of 0.66 to
make the IMFs consistent; see Section~\ref{sec:sfrcalculation}) to
obtain a typical intrinsic H$\alpha$ luminosity, $L_{\rm H\alpha,
int}$, for each bin, i.e.\
\begin{equation}
\left(\frac{L_{\rm H\alpha, int}}{\rm erg~s^{-1}}\right) =
8.35\times10^{40} \left(\frac{\Psi_{24}}{M_{\odot}~{\rm yr}^{-1}}\right).
\label{eq:halphasfr}
\end{equation}
As galaxies are all at a known redshift, these luminosities are 
converted to the observed line fluxes by assuming that a single value
of extinction $A_{\rm H\alpha}$ can be applied to galaxies in each bin,
i.e.\
\begin{equation}
A_{\rm H\alpha} = 2.5~{\rm log}_{10}\left(\frac{L_{\rm H\alpha, int}}{4\pi
  d_{\rm L}^{2} S_{\rm H\alpha}}\right),
\label{eq:halphaextinctionluminosity}
\end{equation}
where $d_{\rm L}$ is the luminosity-distance at $z=0.845$.  Combining
these equations, we obtain
\begin{equation}
A_{\rm H\alpha} = 2.5~{\rm
  log}_{10}\left[\left(\frac{8.35\times10^{40}}{4\pi d_{\rm
  L}^{2}}\right)\left( \frac{\Psi_{24}/M_{\odot}~{\rm yr}^{-1}}{S_{\rm
  H\alpha}/{\rm erg~s^{-1}~cm^{-2}}}\right)\right].
\label{eq:halphaextinction}
\end{equation}
We calculate an extinction from each set of binned data using
equation~(\ref{eq:halphaextinction}), and the results are shown on
Fig.~\ref{fig:halphasfrextinction}(a).  There is an indication of an
increase in $A_{\rm H\alpha}$ with observed H$\alpha$ luminosity; the
best fit to the data is
\begin{equation}
A_{\rm H\alpha} = (-19.46\pm6.77) + (0.50\pm0.16)~{\rm
    log}_{10}\left(\frac{L_{\rm H\alpha, obs}}{{\rm erg~s^{-1}}}\right), 
\label{eq:datafit}
\end{equation}
i.e.\ a $3\sigma$ detection of a variation in extinction with
observed H$\alpha$ luminosity.  However, if there is a dependence on
luminosity (and by implication, on SFR activity) then relating
extinction to observed H$\alpha$ luminosity may obscure some of the
effects.  In order to test this, we show the dependence of extinction
on SFR in Fig.~\ref{fig:halphasfrextinction}(b).  The dependence of
extinction on SFR is more prominent, with a best fit to the data of
$A_{\rm H\alpha} = (0.73\pm0.15) + (0.44\pm0.11)~{\rm
log}_{10}(\Psi_{24}/M_{\odot}~{\rm yr}^{-1})$, i.e.\ a $4\sigma$
detection of SFR-dependent extinction.

Previous studies \citep[e.g.][]{Wang96, Hopkins01, Sullivan01,
Dopita02, Berta03, PerezGonzalez03, Buat05, Schmitt06, Caputi08} have
found a correlation between extinction and either luminosity or SFR,
with the implication being that more actively star-forming regions
within a galaxy are also dustier.  \citet{Hopkins01} derived a
SFR-dependent reddening relationship, based upon the observed Balmer
decrement and far-IR luminosity of a sample of local star-forming
galaxies \citep{Wang96}.  We repeat the derivation from Section~3 of
\citet{Hopkins01} in order to obtain a general expression for their
relationship between SFR and extinction,
\begin{equation}
A_{\rm H\alpha} = \frac{2.5 k_{\rm H\alpha}}{k_{\rm H\beta} - k_{\rm
    H\alpha}} {\rm log}_{10}\left[\frac{0.797~{\rm
    log}_{10}\left(\eta \Psi\right)+3.83}{2.86}\right],
\label{eq:generalextinction}
\end{equation}
where $k_{\lambda}$ is related to the $E(B-V)$ colour excess by
$k_{\lambda} = A_{\lambda}/E(B-V)$, and $k_{\rm H\alpha}$ and $k_{\rm
H\beta}$ are the values of $k_{\lambda}$ at the wavelengths of
H$\alpha$ and H$\beta$ radiation respectively.  $\Psi$ is the total
SFR of a galaxy in $M_{\odot}$~yr$^{-1}$, and $\eta$ is a conversion
factor appropriate to the chosen IMF (1 for a Salpeter IMF, and 0.66
for the Kroupa IMF used in this work).  The factor of $\eta$ in
equation~(\ref{eq:generalextinction}) cancels an equivalent factor of
$1/\eta$ during the calculation of $\Psi$ from IR luminosity in
Section~\ref{sec:sfrcalculation}, and is otherwise unimportant. The
value of 2.86 represents the intrinsic H$\alpha$/H$\beta$ line flux
ratio \citep[appropriate for Case~B recombination, a temperature of
$10^{4}$~K, and an electron density of
$10^{2}$~cm$^{-3}$;][]{Brocklehurst71}, and the other numerical
factors come from the least-squares fit performed by \citet{Hopkins01}
to the observed Balmer decrement and the logarithmic far-IR luminosity
of local star-forming galaxies.

In order to test the form of the dust attenuation law, we assume that
the numerical factors within the square brackets of
equation~(\ref{eq:generalextinction}) are fixed and correct, and fit
our data for the numerical prefactor, $B \equiv 2.5k_{\rm H\alpha} /
(k_{\rm H\beta}-k_{\rm H\alpha})$, with errors coming from the typical
uncertainty in $A_{\rm H\alpha}$.  We obtain $B=5.8\pm1.0$, or
alternately
\begin{equation}
\frac{k_{\rm H\beta}}{k_{\rm H\alpha}} = \left(1 +
\frac{2.5}{B}\right) = 1.43\pm0.07.
\end{equation}
This is consistent with the value obtained from the \citet{Calzetti00}
dust attenuation law, which predicts a value of $B=6.54$, and gives
$k_{\rm H\beta}/k_{\rm H\alpha} = 1.38$.  We show this result on
Fig.~\ref{fig:halphasfrextinction}(b) to demonstrate the good
agreement with our data.  Values of $B$ obtained from the individual
fields are also consistent with this value (see
Table~\ref{tab:normalisation}).

\citet{Brinchmann04} have tested the two assumptions that we have made
in the derivation of equation~(\ref{eq:generalextinction}), namely
that the intrinsic H$\alpha$/H$\beta$ ratio is always equal to 2.86,
and that the conversion between intrinsic H$\alpha$ luminosity and SFR
that we use is appropriate for all galaxies.  They conclude that while
both assumptions are slightly flawed -- the Case~B ratio depends
slightly upon temperature, while the most metal-rich galaxies produce
a slightly lower H$\alpha$ luminosity for the same SFR -- these
variations are small and approximately cancel each other out; the
combination of the two assumptions is therefore a fairly good
approximation to make.  Variations in the intrinsic Case~B ratio for
different temperatures \citep[between a ratio of 2.72 and
3.03;][]{Brocklehurst71} could only affect our expected value of
$B=6.54$ by up to $\pm0.16$, which is much smaller than the
statistical errors in this study.

\citet{Sullivan01} and \citet{PerezGonzalez03} have also produced
least-squares fits to the correlation between the Balmer decrement and
H$\alpha$-derived SFR from independent sets of galaxies, leading to
slightly different numerical factors within the square brackets of
equation~(\ref{eq:generalextinction}) .  We show the equivalent
versions of equation~(\ref{eq:generalextinction}), calculated from
these fits, on Fig.~\ref{fig:halphasfrextinction}(b) for comparison
with the \citet{Hopkins01} result.  As noted by \citet{Sullivan01},
the slope of their correlation is very similar to that of
\citet{Hopkins01}, although the independent estimation of SFR from IR
observations \citep[carried out by][and in this work]{Hopkins01}, is
preferable to using H$\alpha$-derived values, providing that AGN
contamination has been removed.  We find that the slope of the
\citet{Sullivan01} relationship agrees with our data, but that the
overall normalisation is too high.  In contrast, the
\citet{PerezGonzalez03} correlation does not agree with our results;
they indicate that if highly-obscured objects are removed from their
sample, they find a correlation similar to that of \citet{Sullivan01}.

We note that \citet{Afonso03} have found a significantly steeper
relationship between SFR and H$\alpha$ extinction from a
radio-selected sample than that obtained in this work or by
\citet{Hopkins01}.  However, while \citet{Hopkins01} found a
reasonable correlation between the Balmer decrement and IR luminosity
of their sample, \citet{Afonso03} did not find a good correlation
between the Balmer decrement and radio-derived SFR of their sources,
which may make their derived relationship less reliable.  

Following \citet{Hopkins01} we combine equations~(\ref{eq:halphasfr})
and~(\ref{eq:generalextinction}) to obtain a relationship between the
extinction of a galaxy and its {\it observed} H$\alpha$ luminosity,
\begin{equation}
A_{\rm H\alpha} = 12.8 \times 10^{A_{\rm H\alpha}} - 2.5~{\rm
  log}_{10}\left[7.9\times10^{-42} \left(\frac{L_{\rm H\alpha, obs}}{\rm
  erg~s^{-1}}\right)\right] - 12.0,
\label{eq:numerical}
\end{equation}
where the observed H$\alpha$ luminosity $L_{\rm H\alpha, obs}$ is the
value calculated from the observed H$\alpha$ line flux, after
corrections for N\,{\sc ii} flux contamination, but before any dust
attenuation correction has been applied to the data.
Equation~(\ref{eq:numerical}) cannot be solved analytically; however,
over the luminosity range covered by sources in this study ($1.6\times
10^{41} < L_{\rm H\alpha, obs} < 4\times10^{42}$~erg~s$^{-1}$), we can
approximate it by the linear fit
\begin{equation}
A_{\rm H\alpha} = -21.963 + 0.562~{\rm log}_{10}\left(\frac{L_{\rm H\alpha,
    obs}}{\rm erg~s^{-1}}\right).
\label{eq:approximate}
\end{equation}
This is consistent with our fit to the data from
equation~(\ref{eq:datafit}), as shown on
Fig.~\ref{fig:halphasfrextinction}(a), and is within 5~per~cent of the
numerical solution over the observed luminosity range tested in this
work.  The approximate and numerical solutions are consistent to
within 10~per~cent over the luminosity range of $10^{41} -
10^{44}$~~erg~s$^{-1}$.

\subsection{Dependence of H$\alpha$ extinction on galaxy properties}
\label{sec:halphaextinctionproperties}
We have shown that equation~(\ref{eq:generalextinction}) is
appropriate for our full sample of H$\alpha$ emitters, and accurately
describes the variation in dust extinction for sources with different
SFRs.  In order to test whether there is any variation in dust
attenuation between different types of galaxies, we split the sample
into subsets based upon other properties of the galaxies.  We stack
each subset into different H$\alpha$ flux bins in order to remove the
dependence of extinction on SFR, as the mean SFR of different galaxy
types can vary strongly.  We then repeat the analysis of
Section~\ref{sec:halphaextinction} in order to verify that
equation~(\ref{eq:generalextinction}) accurately describes the
variations in the data, and to calculate a value of $B$ for the
subset.  A summary of the relationships derived for all subsets of
galaxies can be found in Table~\ref{tab:normalisation}.  

\begin{figure*}
  \begin{center}
    \includegraphics[width=0.9\textwidth]{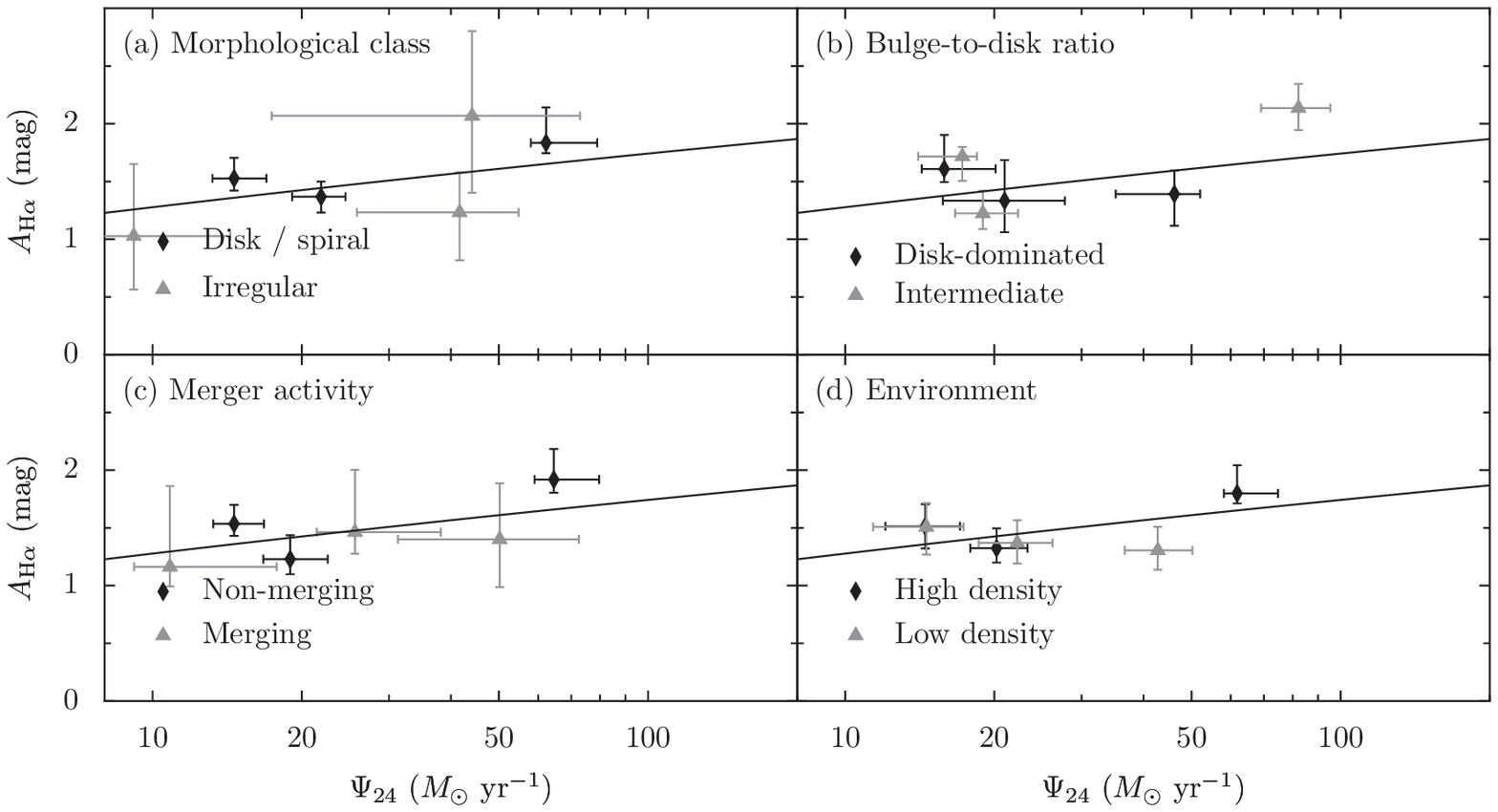}
    \caption{The relationship between H$\alpha$ extinction and SFR,
    for various subsets of the COSMOS data -- the meaning of each
    classification can be found in
    Section~\ref{sec:halphaextinctionproperties}.  The diagonal lines
    show the predicted relationship between $A_{\rm H\alpha}$ and SFR
    from equation~(\ref{eq:generalextinction}) using $B=6.54$, and are
    not fitted to the data.  Values of $B$ obtained through fitting
    equation~(\ref{eq:generalextinction}) can be found in
    Table~\ref{tab:normalisation}, along with errors for each fit.  }
    \label{fig:halphaproperties}
  \end{center}
\end{figure*}
\subsubsection{Morphological type}
\label{sec:morphtype}
A large {\it Hubble Space Telescope} survey of the full 1.8~deg$^{2}$
COSMOS field has been completed \citep{Scoville07HST}, producing
images with $\sim0.1$~arcsec resolution.  Morphological
classifications have been performed on $\sim40$~per~cent of the field
\citep{Scarlata07}, and the majority of the H$\alpha$ emitters have
been assigned a source type (disk/spiral, irregular, early-type) by
\citet{Sobral09Halpha}, based upon this morphological classification,
visual inspection, and colours taken from the Subaru optical data.
There are too few early-type galaxies in the sample for a meaningful
comparison, but the disk/spiral and irregular galaxies are shown in
Fig.~\ref{fig:halphaproperties}(a).

\citet{Scarlata07} have further classified the galaxies with a
disk/spiral morphological type by their bulge-to-disk ratio.  The
number of bulge-dominated galaxies is too small for a meaningful
comparison, but no significant difference is seen between the
intermediate and disk-dominated galaxies in
Fig.~\ref{fig:halphaproperties}(b).

\begin{table}
  \begin{center}
    \caption{The calculated values of normalisation $B$ and error
    $\sigma_{B}$ for equation~(\ref{eq:generalextinction}), broken
    down by the source selection criteria.  The number of galaxies that
    satisfied a particular selection criteria, N, is listed for
    reference -- if N was not large enough for meaningful
    statistics, no normalisation is given.}
    \label{tab:normalisation}
    \begin{tabular}{lccc}
\hline
Sample                                 & N   & $B$ & $\sigma_{B}$\\
\hline
All H$\alpha$ emitters                 & 618 & 5.8 & 1.0\\
COSMOS                                 & 422 & 6.2 & 0.8\\
UDS                                    & 196 & 4.9 & 1.5\\
\hline
COSMOS:\\
disk / spiral morphology       & 347 & 6.9 & 0.7\\
irregular morphology           &  57 & 6.3 & 2.4\\
early-type morphology          &  12 & -   & -  \\
disk / spiral, disk-dominated  & 117 & 6.5 & 1.1\\
disk / spiral, intermediate    & 155 & 7.3 & 0.7\\
disk / spiral, bulge-dominated &  21 & -   & -  \\
merging                        &  92 & 6.0 & 1.9\\
not merging                    & 291 & 6.9 & 0.7\\
low-density                    & 207 & 6.8 & 0.8\\
high-density                   & 207 & 6.3 & 0.9\\
$9.5<$ log$_{10}$(stellar mass/$M_{\odot}$) $<9.8$   &  67 & 1.6 & 1.6\\
$9.8<$ log$_{10}$(stellar mass/$M_{\odot}$) $<10.1$  & 141 & 4.6 & 0.9\\
$10.1<$ log$_{10}$(stellar mass/$M_{\odot}$) $<10.4$ & 180 & 6.6 & 0.6\\
$10.4<$ log$_{10}$(stellar mass/$M_{\odot}$) $<10.7$ & 135 & 7.8 & 0.7\\
$10.7<$ log$_{10}$(stellar mass/$M_{\odot}$) $<11.0$ &  57 & 6.8 & 2.7\\
\hline
Predicted value from \citet{Calzetti00} & & 6.54 & \\
\hline
\end{tabular}
\end{center}
\end{table}

\subsubsection{Merger activity}
\citet{Sobral09Halpha} have visually inspected the galaxies and
classified them according to whether they are undergoing mergers (93
galaxies), or definitely not merging (292 galaxies).  The remaining
galaxies were potential mergers; we have not included these in the
analysis as it is unclear which class they should belong to.  The
merging galaxies and non-merging galaxies are shown on
Fig.~\ref{fig:halphaproperties}(c).

\subsubsection{Environment}
\label{sec:environment}
The angular distances to all nearby sources at comparable redshift
have been calculated for each H$\alpha$ emitter, and converted into a
linear distance to the 10th-nearest neighbour, $r_{10}$ (D.\ Sobral et
al., in preparation).  From these distances, a local projected
surface density can be calculated using $\Sigma = 10/\pi r_{10}^{2}$.
We split the sample in half according to the local density; as shown
in Fig.~\ref{fig:halphaproperties}(d), no significant difference is
seen between the two subsets.

\subsubsection{Stellar mass}
\begin{figure*}
  \begin{center}
    \includegraphics[width=0.9\textwidth]{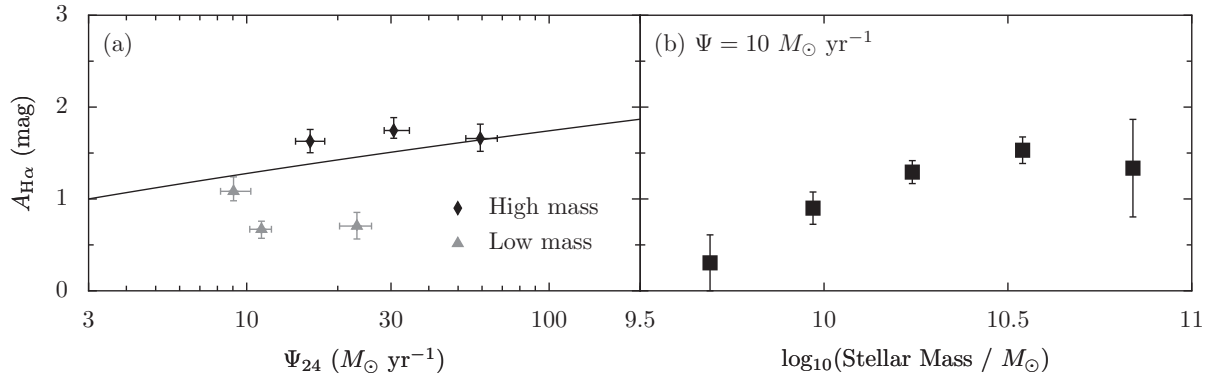}
    \caption{The relationship between H$\alpha$ extinction and SFR for
    data binned by stellar mass.  (a) Data separated into two stellar
    mass bins.  (b) The extinction that would be predicted for a
    galaxy of a given stellar mass, and a SFR of $\Psi =
    10$~$M_{\odot}$~yr$^{-1}$, with the data separated into five mass
    bins.}
    \label{fig:extinctionmass}
  \end{center}
\end{figure*}

Stellar mass estimates are available for all COSMOS and UDS sources
\citep[M.\ Cirasuolo et al., in preparation]{Mobasher07}.  We split
our sample in half by stellar mass, as shown in
Fig.~\ref{fig:extinctionmass}(a).  Two results are apparent from this
plot -- bins with lower stellar mass also have a lower median SFR, and
the lowest stellar mass bin appears to have a decreased normalisation
compared with the higher mass bin.

Previous studies have shown that the stellar mass and SFR of a galaxy
are correlated \citep[e.g.][]{Brinchmann04}, and the first of our
results is consistent with these studies.  In order to investigate the
second result further, we divide our sample into five stellar mass
bins (and three H$\alpha$ flux bins), calculate normalisations for
each sub-sample, and use equation~(\ref{eq:generalextinction}) to
predict the extinction that would be obtained for a galaxy of this
stellar mass, and a SFR of 10~$M_{\odot}$~yr$^{-1}$.  These
extinctions are shown in Fig.~\ref{fig:extinctionmass}(b).  We find
that galaxies with stellar mass above $\sim10^{10.2}$~$M_{\odot}$ have
an approximately constant extinction / SFR relationship (i.e.\ for a
given SFR, the predicted extinction would be the same), but that below
this stellar mass, the typical extinction for a galaxy with given SFR
decreases significantly.  This may be a result of incompleteness in
our sample, based upon the stellar mass distribution of our H$\alpha$
emitters (which we find to peak at $\sim10^{10.2}$~$M_{\odot}$).  We
do however note that \citet{Brinchmann04} have detected a dependence
of H$\alpha$ extinction on stellar mass for galaxies within SDSS,
which they attribute to a correlation between mass and metallicity for
star-forming galaxies.  Their results do not take into account the
dependence of extinction on SFR which we have confirmed in this work,
and so may simply be a consequence of a correlation between SFR and
stellar mass for star-forming galaxies.  We defer further
investigation of this relationship to a separate work (Garn et al.\ in
preparation).

\subsubsection{Summary}
With the exception of the potential stellar mass dependence, we find
no statistically significant difference between the normalisations of
the relationship between extinction and SFR for any of these subsets
of galaxies, as shown in Table~\ref{tab:normalisation} and
Fig.~\ref{fig:halphaproperties}.  This may simply be a consequence of
the limited number of galaxies available -- after splitting the COSMOS
galaxies into two or three classifications, and a further three
H$\alpha$ flux bins, there are only typically a few tens of galaxies
contained within each bin, and any difference would have to be very
large in order to be detectable.  Note that plots of extinction
against observed luminosity (not shown) also indicate that there is no
significant difference between the various subsets, but we have chosen
to illustrate this fact by relating the extinction to
24-$\mu$m-estimated SFR rather than to observed H$\alpha$ luminosity,
so that a direct comparison to the results of \citet{Hopkins01} can be
made.

We conclude that within the uncertainties quoted in
Table~\ref{tab:normalisation}, the relationship between extinction and
SFR that was presented in equation~(\ref{eq:generalextinction}) is
appropriate for use in galaxies with a wide range of morphologies and
properties.

\subsection{UV dust extinction}
\label{sec:uvextinction}
The UV luminosity of a galaxy is also produced directly from young,
high-mass stars, and will be attenuated by interstellar dust.  We
estimate the dust extinction in the rest-frame UV, $A_{\rm UV}$, in a
similar manner to Section~\ref{sec:halphaextinction}.  As before, we
assume that $\Psi_{24}$ is a good proxy for the true total SFR, and
use an IMF-modified version of the \citet{Kennicutt98} calibration to
calculate a typical UV luminosity, $L_{\nu}$, for each bin, through
\begin{equation}
\left(\frac{L_{\nu}}{\rm W~Hz^{-1}}\right) =
4.71\times10^{20}\left(\frac{\Psi_{24}}{M_{\odot}~{\rm yr}^{-1}}\right).
\end{equation}
$A_{\rm UV}$ is then estimated through comparison of $L_{\nu}$ and an
observed flux density, $S_{\nu}$, using 
\begin{equation}
A_{\rm UV} = 2.5~{\rm log}_{10}\left[\frac{L_{\nu}(1+z)}{4\pi d_{\rm
    L}^{2} S_{\nu}}\right],
\label{eq:uvextinctionluminosity}
\end{equation}
where the $1+z$ factor accounts for bandwidth stretching.  

\begin{table}
  \begin{center}
    \caption{The rest-frame wavelength, zero-point correction and
    median galactic extinction for each of the three bands used to
    calculate the UV extinction, along with the calculated conversion
    factor $\gamma$ between nebular and stellar reddening for each
    band from equation~(\ref{eq:uvextinction}), using the stacked data
    (Fig.~\ref{fig:uvsfrextinction}), and from individual sources
    (Fig.~\ref{fig:halphauvratio}).}
    \label{tab:uv}
    \begin{tabular}{lccc}
\hline
 & $u^{*}$ & $B_{J}$ & $g^{+}$\\
\hline
Rest-frame $\lambda$ (\AA)       & 2058.5 & 2417.2 & 2590.6\\
Zero-point correction  & $-0.084$ & 0.189 & $-0.090$\\
Galactic extinction & $-0.086$ & $-0.074$ & $-0.069$\\
\hline
$\gamma$ (stacked data) & $0.50\pm0.07$ & $0.50\pm0.07$ &
 $0.48\pm0.07$\\ 
$\gamma$ (individual sources) & $0.49\pm0.14$ & $0.49\pm0.14$ &
 $0.47\pm0.14$\\ 
\hline
\end{tabular}
\end{center}
\end{table}

\begin{figure}
  \begin{center}
    \includegraphics[width=0.45\textwidth]{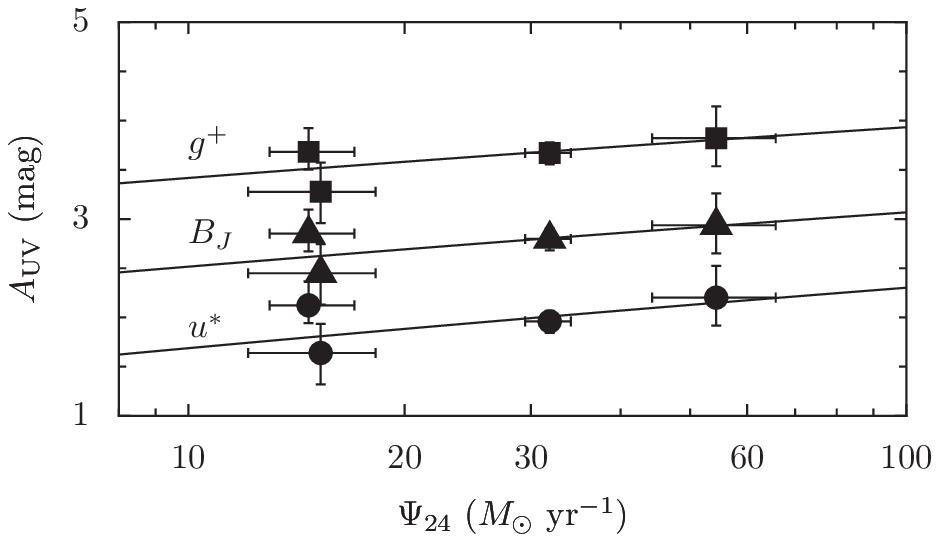}
    \caption{Variation in the median UV extinction with 24-$\mu$m SFR,
    for the HiZELS COSMOS emitters after selection by their H$\alpha$
    flux.  The UV extinction has been calculated from photometry in
    the $u^{*}$ band, $B_{J}$ band (offset by $+1$~mag) and $g^{+}$
    band (offset by $+2$~mag).  The solid lines are fits to
    equation~(\ref{eq:uvextinction}), using the \citet{Calzetti00}
    dust attenuation law to calculate $k_{\rm UV}$, and with values of
    $\gamma$ as listed in Table~\ref{tab:uv}.}
    \label{fig:uvsfrextinction}
  \end{center}
\end{figure}

We have three photometric bands that trace far-UV flux density
($u^{*}$, $B_{J}$ and $g^{+}$), with the rest-frame wavelengths that
they probe listed in Table~\ref{tab:uv}.  We only consider COSMOS
galaxies in this section, as there are currently no accurate
photometric corrections available for the UDS data.  We applied the
photometric zero-point corrections that are recommended by
\citet{Capak07} to the photometry of each source, along with
individual corrections for galactic extinction (with the median
correction factors listed in Table~\ref{tab:uv}).  An appropriate
aperture correction \citep[which applies to each band; see][]{Capak07}
was taken from the PSF-matched catalogue for each source, with a
median correction of $-0.275$~mag.  Fig.~\ref{fig:uvsfrextinction}
shows the measured dependence of UV extinction on SFR for the three
bands -- as before, there is a correlation between the SFR and UV
extinction.

\citet{Calzetti97} found that the dust obscuration of stellar
continuum emission between the wavelengths of the H$\alpha$ and
H$\beta$ lines is less than would be expected from an extrapolation of
the obscuration of the line emission.  They found that the optical
depth of the dust that was obscuring the continuum was about
60~per~cent of the depth of the dust that obscures the line emission.
The inference drawn from this was that massive stars (producing the
H$\alpha$/H$\beta$ emission) form in dusty environments, while the
bulk of the red stellar continuum was a result of much less
dust-obscured stars (presumably older, and lower mass), at a different
location in the galaxy.  They suggest using $E(B-V)_{\rm star} =
(0.44\pm0.03) \times E(B-V)_{\rm gas}$ to calculate the reddening of
the continuum emission.  

The difference of environment should not be true for H$\alpha$ and UV
continuum radiation, which are both predominantly produced by the same
population of young stars; it might therefore be expected that the
correction factor (which we denote as $\gamma$) should converge to 1
for shorter wavelength continuum radiation, where the obscuring dust
will become more like the dust that obscures the H$\alpha$ emission.
Note, however, that the value of $\gamma$ is closely linked to the
shape of the \citet{Calzetti00} dust law, which has been calculated
through comparisons of the SEDs of attenuated galaxies and their
unattenuated counterparts -- different wavelength regions of the dust
attenuation law relate to different regions of a galaxy (i.e.\ the UV
end is physically related to the dusty star-forming regions, while the
long wavelength end will be weighted towards the old stellar
population residing in less dusty regions).  As such, there is a
degeneracy between the wavelength dependence of $\gamma$ and
$k_{\lambda}$ that does not relate to physical properties of dust
within galaxies.  We therefore believe that while the value of
$\gamma$ is useful in calculating UV attenuation from the Balmer
decrement and vice versa, it does not have a direct physical
relevance, as it is closely linked with the assumed form of the dust
law.  If UV extinction at $z=0.84$ is the same as that seen locally,
and described by the \citet{Calzetti00} dust law, we would expect to
find $\gamma\approx0.44$ for UV emission as well as for longer
wavelength continuum emission.

We allow the conversion between nebular and stellar reddening,
$\gamma$, to be a free parameter, and adjust
equation~(\ref{eq:generalextinction}) by this factor to obtain
\begin{equation}
A_{\rm UV} = \frac{2.5 \gamma~k_{\rm UV}}{k_{\rm H\beta} - k_{\rm
H\alpha}} {\rm log}_{10}\left[\frac{0.797~{\rm log}_{10}(\eta \Psi) +
3.83}{2.86}\right],
\label{eq:uvextinction}
\end{equation}
where $k_{\rm UV}$ is the dust attenuation at the rest-frame
wavelength of interest.  We fit for $\gamma$ by assuming that the
\citet{Calzetti00} dust attenuation law accurately estimates $k_{\rm
UV}$ (which we verify through a comparison of the results from
different bands), and show these fits on
Fig.~\ref{fig:uvsfrextinction}.  The values of $\gamma$ are given in
Table~\ref{tab:uv} -- all are larger than 0.44, but are within
$\sim1\sigma$ of the \citet{Calzetti97} result.

\begin{figure}
  \begin{center}
    \includegraphics[width=0.45\textwidth]{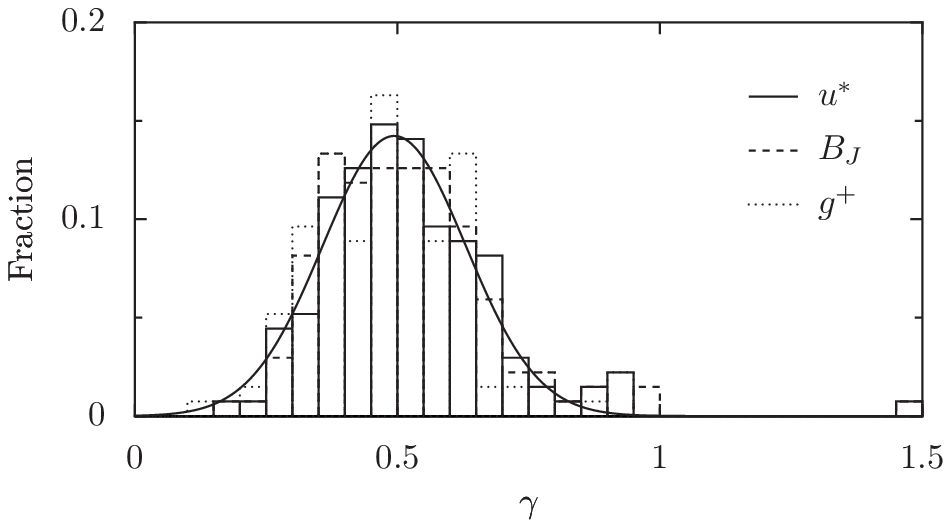}
    \caption{The distribution of $\gamma$ for COSMOS H$\alpha$
    emitters with individual detections at 24-$\mu$m, calculated from
    the $u^{*}$, $B_{J}$ and $g^{+}$ bands.  The Gaussian is a fit to
    the $u$-band data, with ${\bar \gamma} = 0.50$ and
    $\sigma=0.14$.}
    \label{fig:halphauvratio}
  \end{center}
\end{figure}

We calculate the H$\alpha$ and UV extinction for each of the 134
COSMOS H$\alpha$ emitters with a 24-$\mu$m counterpart in the MIPS
catalogue, that were not flagged as AGN.  A strong correlation is seen
between $A_{\rm H\alpha}$ and each of the $A_{\rm UV}$ measurements
(with $\rho=0.70$), and from these extinctions we calculate $\gamma$
on a source-by-source basis.  The distribution of $\gamma$ is shown in
Fig.~\ref{fig:halphauvratio}, and the results given in
Table~\ref{tab:uv} -- in each case the data is consistent with a
Gaussian centred on 0.5, with $\sigma=0.14$.  A two-sample
Kolmogorov-Smirnov test confirms that no significant difference is
seen between the distributions of $\gamma$ in the three bands.  We
therefore conclude that $\gamma = 0.50\pm0.14$, i.e.\ the dust
attenuation of UV emission is approximately half of that which would
be predicted from a simple extrapolation of nebular reddening using
the \citet{Calzetti00} dust attenuation law.  This is in agreement
with the low-redshift result, and implies that the \citet{Calzetti00}
law is broadly applicable to our galaxies.

The principal qualitative difference between the \citet{Calzetti00}
dust attenuation law, derived from local starburst galaxies, and the
individual extinction laws measured from observations of stars within
the Milky Way or the Large Magellanic Cloud is the absence of a `bump'
at around 2175~\AA\, in the rest-frame UV.  There is a conceptual
difference between dust `extinction' by an intervening dust cloud, and
dust `attenuation' from an extended source by a distribution of dust
that is mixed with the stars \citep[see][for more
details]{Calzetti97}.  However, \citet*{Gordon97} have shown that is
it not possible to explain the absence of a 2175~\AA\ dust feature
purely through geometrical effects, implying that the lack of this
feature is intrinsic to the properties of dust grains within local
starburst galaxies.

There are indications that a few high-redshift sources may contain
this dust feature
\citep[e.g.][]{Solorzano04,Srianand08,Eliasdottir09,Noterdaeme09}.  A
2175~\AA\ bump would show up prominently in our data, via a greater
extinction in the $u^{*}$-band data than in the $B_{J}$ and
$g^{+}$ bands.  As no such difference has been found, we
conclude that the dust attenuation law of high-redshift star-forming
galaxies does not require the addition of a UV bump.

\section{Discussion}
\label{sec:discussion}
\subsection{Variation in H$\alpha$ dust extinction with SFR}
H$\alpha$ emission is an excellent tracer to use for cosmic SFR
studies as it is sensitive, can be easily studied over large volumes,
and has a clean selection function.  However, it is necessary to
correct the observed H$\alpha$ luminosity for the effects of dust
extinction before the true SFR can be obtained.

We have shown in Section~\ref{sec:results} that the empirical
relationship between the SFR of local galaxies and their H$\alpha$
extinction is also appropriate for galaxies at $z=0.84$.  We have
further split our sample of galaxies by their morphological type,
bulge-to-disk ratio, merger activity and the local density of their
environments.  No statistically significant difference was seen in the
normalisations of the SFR / H$\alpha$ extinction relationship,
implying that this is a general property of galaxies which applies at
least over the range $0 < z < 0.84$, to spiral and irregular galaxies,
disk-dominated galaxies and galaxies with a moderate bulge component,
to galaxies that are and are not undergoing merger activity, and
galaxies in high and low density environments.

The increase in extinction with higher SFR can be explained through
the fact that star formation takes place within dusty molecular
clouds, and the SFR is related to the gas surface density \citep[the
Schmidt-Kennicutt law;][]{Schmidt59, Kennicutt89} -- more actively
star-forming regions are likely to have dustier environments, where
the H$\alpha$ luminosity produced by young stars will suffer a greater
amount of attenuation.

Equation~(\ref{eq:generalextinction}) allows us to estimate the
typical extinction that is appropriate for galaxies with different
SFRs; the value of $A_{\rm H\alpha} = 1$~mag -- commonly taken to be
the amount of dust extinction to correct for -- is appropriate for
galaxies with a SFR of $\sim3$~$M_{\odot}$~yr$^{-1}$, comparable to
the typical rates of star formation in the local Universe.  The
typical extinction that is appropriate for galaxies with a SFR of 10
and 100~$M_{\odot}$~yr$^{-1}$ is $\sim1.3$ and $\sim1.7$~mag, although
extinctions calculated for individual galaxies may vary considerably
from these values.  \citet{Dopita02} find a range of H$\alpha$
extinctions between $\sim1$ and 3 for galaxies with SFR of
$\sim100$~$M_{\odot}$~yr$^{-1}$, and Fig.~\ref{fig:halphafluxflux}
shows a similar spread of IR-derived H$\alpha$ extinctions for
individual sources.

\subsection{Variation in H$\alpha$ dust extinction between $0<z<0.84$}
Several works \citep[e.g.][]{Tresse07,Villar08} find that the amount
of dust extinction is greater at high redshift than in the local
Universe, with \citet{Villar08} obtaining a mean extinction of $A_{\rm
H\alpha} = 1.48$~mag from a sample of H$\alpha$ emitters at $z=0.84$.
We also find this increase, with a typical extinction value within
$1\sigma$ of 1.5~mag; however, our results from this section and
Section~\ref{sec:halphaextinction} demonstrate that this merely
reflects the greater number of high-SFR sources at higher redshift,
rather than an intrinsic increase in extinction for similar galaxies
at the two redshifts.  We have excluded the possibility of there being
no dependence of H$\alpha$ extinction on SFR at the $4\sigma$ level, a
conclusion that is in agreement with the results of \citet{Choi06} and
\citet{Caputi08}, who find that variations in $V$-band extinction for
different galaxies are correlated with luminosity, rather than with
redshift.

The typical extinction that is measured from a survey that detects
H$\alpha$ emitters at a specific redshift will be dependent on the
detection threshold -- counter-intuitively, a deeper survey will lead
to a {\it lower} measurement of the average extinction, since from
Fig.~\ref{fig:halphasfrextinction}, a lower luminosity threshold also
implies a lower extinction threshold.  We use this result to caution
against drawing conclusions about the typical dust extinction of
galaxies at a given redshift, without taking into account any SFR
dependency.

\subsection{The effect of variable extinction on the
  H$\alpha$ luminosity function and star formation rate density at $z=0.84$} 

\citet{Sobral09Halpha} presented an extinction-corrected H$\alpha$
luminosity function at $z=0.84$ from the HiZELS data used in this
work.  A single extinction of $A_{\rm H\alpha} = 1$~mag was used to
correct all sources, and a Schechter function was fitted to the data,
yielding values of $L^{*}$, $\phi^{*}$ and $\alpha$ which are
presented in column~1 of Table~\ref{tab:schechter}.

We recalculate the H$\alpha$ luminosity function at $z=0.84$ from all
H$\alpha$ emitters, applying the extinction correction from
equation~(\ref{eq:approximate}), and obtain new values of $L^{*}$,
$\phi^{*}$ and $\alpha$, listed in column~2 of
Table~\ref{tab:schechter}.  These results remain consistent with
strong evolution in $L^{*}$ and $\phi^{*}$ over the range $0 < z < 1$,
as found in \citet{Sobral09Halpha}.  The most significant change is an
increase in the characteristic turnover luminosity, $L^{*}$; the
increase in the typical H$\alpha$ extinction leads to a greater number
density of bright emitters, compared with taking $A_{\rm
H\alpha}=1$~mag for all sources.

\begin{table}
  \begin{center}
  \caption{Schechter function parameters $L^{*}$, $\phi^{*}$ and
  $\alpha$ for the $z=0.84$ H$\alpha$ luminosity function, along with
  the average SFRD (in $M_{\odot}$~yr$^{-1}$~Mpc$^{-3}$) obtained by
  integrating the luminosity function down to the HiZELS survey limit
  of $L_{\rm H\alpha}=10^{41.5}$~erg~s$^{-1}$, or across the total
  luminosity range.  The H$\alpha$ extinction assumed when calculating
  the luminosity function is either `1~mag' \citep{Sobral09Halpha} or
  `Variable' (the results from this work), and the AGN correction
  applied when calculating the SFRD either uses all H$\alpha$
  emitters to calculate the luminosity function and then assumes a
  15~per~cent contamination fraction, or only uses those emitters
  which were not identified as AGN.}
  \label{tab:schechter}
    \begin{tabular}{lccc}
\hline
Corrections:\\
Extinction & 1~mag & Variable & Variable\\
AGN  & 15~per~cent & 15~per~cent & AGN excluded\\
\hline
Parameter:\\
log$_{10}(L^{*}/{\rm erg~s^{-1}})$ & $42.26\pm0.05$ & $42.50\pm0.04$ & $42.44\pm0.05$\\
log$_{10}(\phi^{*}/{\rm Mpc^{-3}})$ & $-1.92\pm0.10$ & $-2.05\pm0.09$ & $-1.99\pm0.10$\\
$\alpha$ & $-1.65\pm0.15$ & $-1.60\pm0.10$ & $-1.40\pm0.20$\\
\hline
SFR density:\\
$>10^{41.5}$~erg~s$^{-1}$ & $0.15\pm0.02$ & $0.24\pm0.05$ & $0.15\pm0.01$\\
Total & $0.37\pm0.18$ & $0.41\pm0.16$ & $0.32\pm0.12$\\
\hline
\end{tabular}
\end{center}
\end{table}

\citet{Sobral09Halpha} calculated the average star formation rate
density (SFRD) at $z=0.84$, by integrating the luminosity function and
applying an average 15~per~cent correction for AGN contamination.  If
we use a variable H$\alpha$ extinction correction, and simply apply a
15~per~cent AGN contamination correction, we obtain a significantly
larger value for the SFRD than that found from using 1~mag of
extinction (column~2 of Table~\ref{tab:schechter}).  However, we can
improve upon this estimate by recalculating the H$\alpha$ luminosity
function, excluding those sources which were classified as potential
AGN (column~3 of Table~\ref{tab:schechter}).  There is a flattening
seen towards the faint end of the luminosity function, with $\alpha$
consistent with a value of $-1.4$.  We integrate this luminosity
function down to the limit of the HiZELS survey, $L_{\rm
H\alpha}=10^{41.5}$~erg~s$^{-1}$, and obtain a SFRD of
$0.15\pm0.01$~$M_{\odot}$~yr$^{-1}$~Mpc$^{-3}$, essentially identical
to the value obtained by \citet{Sobral09Halpha} of
$0.15\pm0.02$~$M_{\odot}$~yr$^{-1}$~Mpc$^{-3}$.  By integrating the
entire luminosity function, we obtain a SFRD of
$0.32\pm0.12$~$M_{\odot}$~yr$^{-1}$~Mpc$^{-3}$, compared with the
value from \citet{Sobral09Halpha} of
$0.37\pm0.18$~$M_{\odot}$~yr$^{-1}$~Mpc$^{-3}$.  The combination of an
improved AGN identification method and variable dust attenuation only
has a moderate effect on the calculated SFRD at $z=0.84$, although may
be more important at higher redshift.

\section{Conclusions}
\label{sec:conclusions}
We have used a well-defined sample of H$\alpha$-emitting galaxies at
$z=0.84$ to study the link between star formation and dust extinction.
The HiZELS sample consisted of 694 H$\alpha$-emitting galaxies within
the redshift range $z=0.845\pm0.021$ in the well-studied COSMOS and
UDS fields, with additional information available in multiple UV,
optical and IR bands, along with ancillary radio and X-ray data.

In order to focus on star-forming galaxies, it was necessary to remove
AGN contaminants from our sample.  We performed this removal through a
number of multi-wavelength classification methods; emission-line
diagnostics, deviations away from the IR / radio correlation, X-ray
counterparts, mid-IR and mid- to far-IR shapes and full template
fitting were all used to identify potential AGN.  We estimate an AGN
contamination rate of between 5 and 11~per~cent for H$\alpha$-selected
emission-line sources at $z=0.84$.  No single diagnostic identified
more than half of the AGN contaminants, with SED template fitting
being the most successful; we suggest that using a combination of
template-fitting and colour-colour diagnostics will maximise the
success rate of AGN identification in H$\alpha$ surveys.

In order to constrain the average IR luminosity of H$\alpha$ emitters
at $z=0.84$, we have binned our sample by observed H$\alpha$ flux, and
used the technique of source stacking to calculate typical 24-$\mu$m
flux densities for the H$\alpha$ emitters.  Several previous studies
have found that dust extinction correlates with either luminosity or
SFR, and we looked for a relationship between the SFR of a galaxy
and its dust attenuation.  We found that individual extinctions vary
significantly between galaxies, but confirm that this correlation
holds for the average properties of our sample; galaxies with higher
SFR also undergo a greater amount of extinction.  

We have presented an approximate method of calculating the typical
H$\alpha$ extinction of a galaxy, based solely upon the observed
H$\alpha$ luminosity; $A_{\rm H\alpha} = -21.963 + 0.562~{\rm
log}_{10}(L_{\rm H\alpha, obs}/{\rm erg~s^{-1}})$, which is within
10~per~cent of the full numerical solution over the luminosity range
$10^{41} - 10^{44}$~erg~s$^{-1}$.  This is found to describe the
properties of galaxies with disk, spiral and irregular morphologies, a
range of bulge-to-disk ratios, galaxies in high- and low-density
environments, and both merging and non-merging galaxies, although a
potential deviation in this relationship may be seen for galaxies with
low stellar masses.  We emphasize that this is only an average
relationship, and that the extinction measured within individual
galaxies can vary away from this relationship by up to $\sim2$~mag.

We have calculated the far-UV extinction of galaxies in a similar
manner to the H$\alpha$ extinction, and found a qualitatively similar
correlation between this extinction and the SFR.  We find that the UV
dust obscuration is about half of that expected from an extrapolation
of the \citet{Calzetti00} dust attenuation law.  By calculating UV
extinctions from flux densities measured in three different bands, we
show that, similar to the local Universe, there is no evidence for a
2175~\AA\ UV bump in the dust attenuation of high-redshift
star-forming galaxies.

The relationship between dust extinction and SFR at $z=0.84$ is the
same as is seen at $z\sim0$; we conclude that there is no evidence for
a change in the dust properties of galaxies over this redshift range,
and that any average increase in the extinction of galaxies at higher
redshift is due to the overall increase in star formation activity out
to $z\sim2$.  Using our relationship between observed H$\alpha$
luminosity and extinction, we recalculate the H$\alpha$ luminosity
function for galaxies at $z=0.84$.  We find an increase in the
characteristic luminosity compared with the value obtained by assuming
a single value of extinction is applicable to all sources, implying a
greater number density of bright H$\alpha$ emitters than previous
estimates at $z=0.84$.

\section*{Acknowledgements}
TSG and PNB are grateful for support from the Leverhulme Trust.  DS
would like to thank the Funda{\c c}{\~a}o para a Ci{\^e}ncia e
Tecnologia (FCT) for the doctoral fellowship SFRH/BD/36628/2007.  JEG
and IRS acknowledge support from STFC.  RJM and JSD acknowledge the
support of the Royal Society through a University Research Fellowship
and a Wolfson Research Merit award respectively.  DF thanks STFC for
support via an Advanced Fellowship.

\setlength{\labelwidth}{0pt}
\bibliography{./References}
\bibliographystyle{/home/tsg/data/Documents/Papers/mn2e}
\label{lastpage}

\end{document}